\PassOptionsToPackage{table}{xcolor}
\documentclass[acmsmall,nonacm]{acmart}

\usepackage[normalem]{ulem} 
\usepackage[safe]{tipa}
\usepackage{forest}
\usepackage{booktabs} 
\usepackage{subcaption} 
\usepackage{tcolorbox}
\usepackage{syntax}
\usepackage{listings}
\usepackage{svg}
\usepackage{comment}
\usepackage{mathpartir}

\makeatletter
\let\old@lstKV@SwitchCases\lstKV@SwitchCases
\def\lstKV@SwitchCases#1#2#3{}
\makeatother
\usepackage{lstlinebgrd}
\makeatletter
\let\lstKV@SwitchCases\old@lstKV@SwitchCases

\lst@Key{numbers}{none}{%
   \def\lst@PlaceNumber{\lst@linebgrd}%
   \lstKV@SwitchCases{#1}%
   {none:\\%
    left:\def\lst@PlaceNumber{\llap{\normalfont
               \lst@numberstyle{\thelstnumber}\kern\lst@numbersep}\lst@linebgrd}\\%
    right:\def\lst@PlaceNumber{\rlap{\normalfont
               \kern\linewidth \kern\lst@numbersep
               \lst@numberstyle{\thelstnumber}}\lst@linebgrd}%
   }{\PackageError{Listings}{Numbers #1 unknown}\@ehc}}
\makeatother

\usepackage{hyperref}
\usepackage[capitalise]{cleveref}
\usepackage{titletoc}
\usepackage{multirow}
\usepackage{makecell}
\usepackage{custommacro}
\usetikzlibrary{decorations.pathreplacing}

\newcommand{\node}[0]{\textsf{node}}

\makeatletter
\@addtoreset{equation}{enumi}
\makeatother

\Crefname{figure}{Fig.}{Figs.}

\newcommand{\umang}[1]{}

\makeatletter
\newcommand{\bigplus}{%
\DOTSB\mathop{\mathpalette\mattos@bigplus\relax}\slimits@
}
\newcommand\mattos@bigplus[2]{%
\vcenter{\hbox{%
 \sbox\z@{$#1\sum$}%
 \resizebox{!}{0.9\dimexpr\ht\z@+\dp\z@}{\raisebox{\depth}{$\m@th#1+$}}%
}}%
\vphantom{\sum}%
}

\theoremstyle{remark}

  \author{Karuna Grewal}
\email{kgrewal@cs.cornell.edu}
\affiliation{%
  \institution{Cornell University}
  \country{USA}
}
\author{P. Brighten Godfrey}
\affiliation{%
  \institution{University of Illinois
Urbana-Champaign}
  \country{USA}
}
\email{pbg@illinois.edu}
\author{Justin Hsu}
\affiliation{%
  \institution{Cornell University}
  \country{USA}
}
\email{justin@cs.cornell.edu}

\author{Umang Mathur}
\affiliation{%
  \institution{National University of Singapore}
  \city{}
  \country{Singapore}
}
\email{umathur@nus.edu.sg}
 \newif\ifappendix
 \appendixfalse
 
  \newcommand{\appref}[1]{%
  \ifappendix
    \S\ref{#1}
  \else
  \fi
}
\begin{document}

\title{SafePar: Monitoring Asynchrony in Microservices}


\begin{abstract}
Modern cloud applications are built from loosely-coupled microservices that
coordinate through well-defined APIs to service user requests. A single API
request often triggers multiple downstream API calls, some executed sequentially
and others spawned asynchronously in parallel. To certify safe and secure
inter-service interactions in such applications, security and compliance teams
must enforce policies not only over nested call/return structure, but also over
the parallel structure of an execution: which calls may run concurrently, how
many parallel branches can be spawned, and what combination of branch outcomes
are allowed.  However, existing runtime enforcement mechanisms typically model
executions as sequential or purely nested traces, and cannot capture the
parallel structure introduced by asynchronous API calls. Furthermore, since
application implementations may not be accessible to security and compliance
teams, the policy enforcement mechanism should be decoupled from the service
implementation.

We introduce SafePar, a specification and monitoring framework for policies over
concurrent microservice executions. A SafePar policy constrains both the order
of API calls and their series-parallel structure. To support seamless
deployments, each policy is compiled into a series-parallel visibly pushdown
automaton, a new model of computation we propose in this work, that drives a
distributed runtime monitor implemented on top of the servicemesh layer. Our
technique is blackbox and non-invasive: it requires no access or changes to the
service implementation. Our experiments show that SafePar enforces rich concurrency-aware policies while incurring only millisecond-scale latency overhead.
\end{abstract}

 \maketitle

 \section{Introduction}

Microservice architecture \cite{microservice} is a widely used design paradigm
for building cloud-native applications. In this architecture, an application is
decomposed into loosely coupled components called microservices, which expose
their functionalities over well-defined API endpoints for other services to
consume. This decomposition enables microservices to be owned, developed, and deployed by independent teams.

This modular design also benefits the application's performance and availability because deployment team can now scale up only the services under load, or replicate only necessary services like backend data stores, instead of scaling up all the components of the application, in contrast to vanilla monolithic architectures. 

However, microservice applications are difficult to develop correctly:
a local change that appears correct from one team's perspective can silently
violate an application-wide safety property, especially when different services
rely on implicit or outdated assumptions about each other's API contracts. Prior
studies \cite{bugmicroservice} report that a substantial share of production in
faults in microservice systems arise from cross-service interactions, and are
difficult to catch with simple unit or and integration tests.
Furthermore, conventional bugs due to logic errors and improper error handling
are still present, and the microservice setting amplifies their impact with its
distributed communication patterns and independently evolving services that make
violations harder to anticipate and debug.

\subsection{Enforcing safety properties for microservices}

In a microservice application, an API call may trigger a tree of API interactions over protocols like HTTP or gRPC. Therefore, realistic safety properties require simultaneously reasoning over  multiple APIs calls in the application's runtime trace rather than a simple single API call in isolation.
For example, a GDPR  regulation \cite{gdpr} may require a database write API to occur only after an encrypt API for a European region user; a CI/CD pipeline may require a software release API only after a testing API and vulnerability scan API succeed. Despite teams not having access to the application's implementation, they want to enforce such safety properties because failure to enforce then can lead to mishandling of sensitive data and loss of client trust.

Policies on microservice communication patterns are a useful way for security or
compliance teams to specify and enforce fine-grained aspects of the application
runtime behavior (i.e., inter-service communication) without peeking into
service code. For instance, the recent ICLR 2026 incident that
leaked reviewer identities was caused by an internal admin OpenReview API being
publicly reachable due to deployment misconfiguration~\cite{iclr}. 
To ensure that policy enforcement aligns with the constraints of deployability,
the online runtime monitoring and enforcement framework should satisfy
the following natural requirements:

\begin{description}
\item[\rm \emph{Non-invasive}.] The monitor should enforce policies 
without requiring changes to the service code which might not be available.

\item[\rm \emph{Distributed and streaming.}] 
The monitoring framework should not require a centralized observer 
that sees the entire execution trace in a serialized manner, since
a centralized observer introduces significant overhead. 
Instead, the framework should allow for the monitor to be \emph{distributed}
across services such that it 
processes API calls/responses as they occur---at their source of origin---and
works in a streaming manner,
i.e., storing only a small amount of monitoring information that gets updated 
as new events in the execution are observed.

\item[\rm \emph{Structure aware.}] 
Many meaningful policies express constraints on the structure of the execution trace,
for example, order of API calls, nesting structure of invocations of other APIs, etc.
The monitoring framework should be flexible and expressive so as to expose
the structure of the execution allowing for precise policy enforcement.
\end{description}

\paragraph{\textbf{Existing work and limitation.}}
Recent work \cite{grewalhotnets,safetree} 
shows how the above desiderata can be met for microservice applications that are
\emph{synchronous},
where a service issuing an API request 
waits for its callee's response before proceeding.
\citet{grewalhotnets} presented a policy language for this setting, where
safety properties could be expressed as regular expression-styled policies over API call order. 
In more recent work, SafeTree \cite{safetree} adds declarative constructs for specifying valid tree and parent-child nesting. Both frameworks enforce these structural policies in a non-invasive, distributed, and incremental manner 
using runtime monitors based on appropriate models of computation models, namely word automata and nested word automata respectively. These approaches are well-suited for executions that can be represented as nested words, where a parent invokes a child and then waits for that child to return before continuing.

Unfortunately, policies that only express vanilla temporal and nesting constraints
are not always sufficient to capture meaningful behaviors in more realistic microservice applications.
Indeed, we are interested in modern microservice-based applications that
make API calls \emph{asynchronously}, allowing them to reduce 
latency and provide higher availability. 
For instance, after authenticating a request, a frontend may issue concurrent
read operations on multiple confidential files,
or a service may concurrently query multiple replicas to improve availability.
Indeed, in modern microservice architectures, services invoke several APIs in parallel, 
each of which performs its (nested) computation independently, 
and then collect all of their responses before continuing. 
Such a fork-join structured parallelism, in turn, means that
correctness requirements may require one to  talk about concurrent sibling branches. For example, a deployment team may want a request sent to multiple backend replicas to continue only if at least one replica returns a valid response, or a security team may want every branch that accesses confidential data to be accompanied with a sibling logging branch. 
Prior work on nested word based monitoring framework cannot express and enforce
such policies because they represent execution sequentially and cannot model
parallel structure in asynchronous traces.

In this paper, we seek to answer the question: \textbf{\textit{ ``How can we express and non-invasively enforce safety properties over structured parallel executions of blackbox microservice applications with asynchronous APIs?''}
} Answering this question requires addressing all the monitoring desiderata from first principles;
the last two are especially challenging and require careful consideration:

\noindent
\paragraph*{\textbf{Challenge 1: need for a series-parallel structure with
matching call-returns.}} The safety properties in the microservice setting
involving asynchrony require us to model executions using structures that
capture two key details about the execution tree: (a) the series-parallel
structure of sub-executions of API calls, and (b) the well-matched and nested
API call/return structure, \textit{i.e.,} each API call has a corresponding
return and  children API's execution is enclosed between the scope of its
parent's API call and return.

\paragraph*{\textbf{Challenge 2: need for an computation model over the above structure with monitoring-amenable semantics.}}
A distributed runtime monitor in the microservice setting observes an execution
one event at a time at its source of origin. This means that the underlying
computation model for such monitors must be able to support \emph{incremental}
input, rather than requiring the entire trace upfront. 
Unfortunately, existing monitor models (most of which are automata
  models) for recognizing series-parallel structures, like graphs, posets, and
  pomsets, are not expressive enough for reasoning about the class of behaviors (concurrency together with well-nesting) that arise in our setting.

\subsection{Our solution}

In this work, we develop a monitoring framework for asynchronous microservice
applications consisting of three parts:
(a) a policy language for expressing microservice communication 
patterns involving asynchronous API calls, 
(b) an automaton model that  incrementally reads 
the events in a series-parallel trace, 
(c) a non-invasive online distributed runtime monitor, atop
servicemesh~\citep{ashok21servicemeshes,servicemesh}, for enforcing policies in our language.
We elaborate on each component below.

\paragraph*{\textbf{Series-parallel nested words policy language for asynchronous microservice behaviors.}}

To model microservice traces with asynchronous API calls, 
we define  \emph{series-parallel nested word} (SPNW) structures.
Conceptually, SPNW structures (or SPNWs for short) extend nested words \cite{nestedwords} 
with a parallel composition operator.
Such structures can then naturally model multiple concurrently executing API calls
made by a parent API, along with the usual constructs of sequential composition
and nested calls.

\umang{This needs to change now. We are no longer pitching SPNWEs as our policy language. We are now pitching the $1$-unambiguous thingy.}
We next introduce SafePar, whose policy language is based on our notion of \emph{series-parallel nested word expressions} (SPNWE). 
SafePar can express properties like:  
(a) A call to API \textsf{A} should be nested inside the scope induced by a call to API \textsf{B}, 
(b) The execution of API \textsf{A} should happen after that of API \textsf{B}, and 
(c) Collective properties over concurrent executions, 
for example: in a group of concurrent execution branches, 
at least/at most/exactly $n$ branches should satisfy property $\varphi$.

\paragraph*{\textbf{A visibly pushdown automaton extension for policy enforcement.}}
We next design an automata model amenable to monitor SPNWs against the
our class of policies.
More concretely, we introduce a novel \emph{series-parallel visibly pushdown
automaton} (SP-VPA) model for accepting SPNW (i.e., nested words with structured
parallelism) by extending the standard visibly pushdown automaton (VPA)
\cite{vpa} model. 
A key ingredient in making this model amenable to our monitoring setup
is the careful design of its semantics. 
SP-VPA transitions on incrementally reading one API call/return 
symbol at a time as it is generated during the application's execution, 
instead of the entire service tree's series-parallel structure upfront,
as required by existing automata models. 
SP-VPA preserves the standard VPA's stack-based semantics for 
sequentially composed nested API calls/returns.  
However, on observing an asynchronous API calls, the automaton spawns into 
one sub-automaton per API call; each sub-automaton incrementally 
processes its assigned concurrent execution; finally all 
the sub-automata join into one automaton after all the APIs return.

\paragraph*{\textbf{Servicemesh based distributed runtime monitor implementation.}}
A service mesh deployment~\citep{ashok21servicemeshes,istio} is suitable for our blackbox monitoring goal because in this layer a  service is co-located with a servicemesh layer's sidecar proxy 
that runs our monitor and observes and controls 
all the incoming/outgoing traffic for its service. 
Following the key idea of carrying automaton configuration in HTTP headers used by SafeTree for VPA-based monitoring, we implement SafePar as a  SP-VPA-based monitor. SafePar's main extension is support for asynchronous calls: concurrent branches are monitored independently and then the caller's proxy join their returned states to compute the next state. 

\paragraph*{\textbf{Contributions.}}
We summarize our technical contributions as follows:
 \begin{enumerate}
 \item a notion of series-parallel nested words (SPNW) to model microservice
   traces with \async API calls and a regular expression-style policy language over SPNW (in \cref{sec:syntax}),
  \item case studies to demonstrate the expressiveness of our policy language (in \cref{sec:casestudies}),
 \item a visibly pushdown automaton (SP-VPA) model to recognize SPNW and
   operational semantics to incrementally process events along with a sound
   compilation from our policies into SP-VPA (in \cref{sec:semantics}),
 \item a SP-VPA-based distributed monitor implementation atop servicemesh layer and its evaluation (in \cref{sec:impl}, \cref{sec:eval}). 
 \end{enumerate}
We survey related work (in \cref{sec:rw}) and conclude with some future directions (in \cref{sec:conclusion}).


\section{Technical Overview}
\label{sec:overview}

This section motivates SafePar through an example and gives an informal tour of its SafePar specification and enforcement.
\umang{SafePar appears twice.}
 The example illustrates the need for policies over asynchronous microservice 
 executions that are sufficiently rich to allow simultaneously expressing
 call-return nesting, sequential order, 
 and collective behavior of independently executing concurrent branches.

\subsection{Example: data administration console}
Consider a data administration console for managing production data. The console lets an operator perform privileged operations like permanently deleting a production database. This functionality is implemented using several APIs, each implemented by different services: \APIname{delete} API, which carries out the deletion; \APIname{log} API, which records a delete request; \APIname{review} API, which sends a request to a reviewer; and \APIname{approve} API, which records a reviewer's approval.

\newpar{Deletion safety property}
Permanently deletion of a production database is often deemed a high-risk operation.
Security teams in larger organization, therefore, often require that every \APIname{delete} request 
should be logged (through a call to an explicity \APIname{log} API). 
In fact, such teams impose higher scrutiny, for example
by additionally requiring that, before the deletion proceeds, 
at least two of the several concurrently requested \APIname{review} branches should be approved. 
A reviewer typically processes an approval by calling \APIname{approve}. 
Review branches that do not approve can nonetheless co-exist,
provided that a minimum number, say at least two, concurrent approvals are obtained.

\begin{wrapfigure}{R}{0.56\linewidth}
\centering
\resizebox{\linewidth}{!}{%
\begin{tikzpicture}[
  >=Stealth,
  syncarr/.style={-{Stealth[length=6pt]}, thick, namecol!85!black},
  asyncarr/.style={-{Stealth[length=6pt]}, thick, pwblue!80!black},
  elbl/.style={font=\scriptsize\sffamily\itshape, fill=white, inner sep=1.4pt},
  stepnum/.style={circle, draw=badred, fill=badred!8, text=badred, inner sep=0pt,
                  minimum size=3.2mm, font=\tiny\bfseries},
]

\node[pev] (delete)   at (0, 3.2)     {\APIname{delete}$_0$};
\node[pev] (log)      at (-4, 1.7)    {\APIname{log}$_0$};
\node[pev] (review1)  at (-1.3, 1.7)  {\APIname{review}$_1$};
\node[pev] (review2)  at (1.3, 1.7)   {\APIname{review}$_2$};
\node[pev] (review3)  at (3.9, 1.7)   {\APIname{review}$_3$};
\node[pev] (approve1) at (-1.3, 0.2)  {\APIname{approve}$_1$};
\node[pev] (approve2) at (1.3, 0.2)   {\APIname{approve}$_2$};

\begin{scope}[on background layer]
  \node[fit=(review1)(review2)(review3)(approve1)(approve2),
        draw=pwblue!45, rounded corners=6pt, fill=pwblue!6, inner sep=7pt] (parblock) {};
\end{scope}

\draw[syncarr]  (delete.south) -- node[elbl, text=namecol!85!black, pos=.55] {\textsf{sync}} (log.north);
\draw[asyncarr] (delete.south) -- node[elbl, text=pwblue!80!black, pos=.6, sloped] {\textsf{async}} (review1.north);
\draw[asyncarr] (delete.south) -- node[elbl, text=pwblue!80!black, pos=.6, sloped] {\textsf{async}} (review2.north);
\draw[asyncarr] (delete.south) -- node[elbl, text=pwblue!80!black, pos=.6, sloped] {\textsf{async}} (review3.north);

\draw[syncarr] (review1.south) -- node[elbl, text=namecol!85!black, pos=.5, left=1pt] {\textsf{sync}} (approve1.north);
\draw[syncarr] (review2.south) -- node[elbl, text=namecol!85!black, pos=.5, left=1pt] {\textsf{sync}} (approve2.north);

\draw[pwblue!70, line width=.7pt, decorate,
      decoration={brace, amplitude=5pt, mirror}]
  (parblock.north west) -- (parblock.south west)
  node[midway, xshift=-6mm, font=\scriptsize\sffamily, text=pwblue!70!black,
       rotate=90] {parallel};

\node[stepnum] at ($(log.north east)+(1.2mm,0.8mm)$)      {1};
\node[stepnum] at ($(review1.north east)+(1.2mm,0.8mm)$)  {2};
\node[stepnum] at ($(review2.north east)+(1.2mm,0.8mm)$)  {2};
\node[stepnum] at ($(review3.north east)+(1.2mm,0.8mm)$)  {2};
\node[stepnum] at ($(approve1.north east)+(1.2mm,0.8mm)$) {3};
\node[stepnum] at ($(approve2.north east)+(1.2mm,0.8mm)$) {3};

\end{tikzpicture}
}
\caption{A valid API call tree for the deletion policy. The \APIname{delete} API first calls \APIname{log}, then spawns three concurrent \APIname{review} requests, two of which invoke \APIname{approve}.}
\label{fig:deletetree}
\end{wrapfigure}
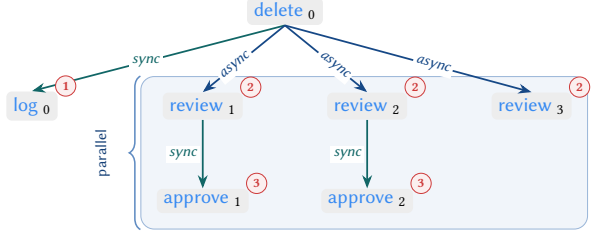

Consider a valid execution (shown in \cref{fig:deletetree}) 
that begins with a call to \APIname{delete}. 
The \APIname{delete} operation first logs the request by synchronously 
calling \APIname{log} and waits for it to return. 
Then, \APIname{delete} sends asynchronous requests to three 
reviewers by concurrently calling \APIname{review} operations. 
In the first two branches, the reviewer approves 
the request by synchronously calling \APIname{approve};
this is followed by
a return from \APIname{approve} API, and then
a return from the respective \APIname{review} API in both these calls to reviewer. 
In the third branch, however, the call to  \APIname{review} returns without an approval. 
The execution of the three review branches is recorded independently.  
Once all three review branches complete, \APIname{delete} aggregates their responses, resumes its execution and eventually returns. 

We are interested in the design of streaming monitors,
and such a monitor cannot see the tree structure induced by the execution in \cref{fig:deletetree} upfront. 
It must instead observe events as they occur. 
Since the three review branches execute concurrently, the monitor may observe their events in different orders. 
Let the call and return events of some API be denoted by its initials $\APIcall{A}$ and $\APIret{A}$, respectively. 
To distinguish which branch produced each event, we annotate the call-return events with a subscript. For example, events in the first review branch are annotated by $1$. Branch $0$ contains the root $\APIname{delete}$ and its synchronous $\APIname{log}$ child, and branches $1, 2, 3$ are the three concurrent review branches. Two possible linearizations that the monitor may observe are:
\begin{align*}
\begin{array}{c}
{\APIcall{D}}_0\ {\APIcall{L}}_0\ {\APIret{L}}_0\ {\APIcall{R}}_1\ {\APIcall{A}}_1\ {\APIcall{R}}_2\ {\APIcall{A}}_2\ 
{\APIret{A}}_1\ {\APIret{A}}_2\ {\APIret{R}}_1\ {\APIret{R}}_2\ {\APIcall{R}}_3\ {\APIret{R}}_3\ {\APIret{D}}_0,\\
{\APIcall{D}}_0\ {\APIcall{L}}_0\ {\APIret{L}}_0\ {\APIcall{R}}_1\ {\APIcall{A}}_1\ {\APIcall{R}}_2\ {\APIcall{A}}_2\ 
{\APIret{A}}_2\ {\APIret{A}}_1\ {\APIret{R}}_2\ {\APIret{R}}_1\ {\APIcall{R}}_3\ {\APIret{R}}_3\ {\APIret{D}}_0.
\end{array}
\end{align*}
These linearizations only differ in the relative order of events 
across independent review branches, while 
the relative ordering of events within a branch is  preserved across the two.   
As a result, both linearizations satisfy the `safe deletion' specification 
and agree on the ordering of logging and review phases with at least two approvals. 
Indeed, this is not surprising, given that the security policy outlined by
the `safe deletion' specification
happens to be invariant under  interleavings of events across different branches. 
In fact, it only restricts an execution's sequential, parallel, and 
parent-child nesting structure, rather than the order picked by a global scheduler. 
Reasoning at the granularity of interleavings (or linearizations)
is often unnecessary, given that most safety properties of interest
are invariant under equivalent interleavings. 
Further, in our setting, where an ideal monitor must be distributed, 
observing a consistent global linearization comes at both a performance cost as well as 
engineering cost.
Nonetheless, monitors cannot be completely oblivious of the structure of the execution
and must reason about temporal order, including nesting and parallel structure, of API calls.
We need a happy medium between a word-like representation for online monitoring, 
but structured enough to record call-return nesting and parallelism.

\subsection{Executions as series-parallel nested words}

SafePar models executions as \textit{series-parallel nested words} (SPNWs).
Intuitively, the \textit{series} aspect of such words demarcate the sequential order
between sub-executions, their \textit{parallel} aspect demarcate
sub-executions that happen in parallel,
and their \textit{nesting} aspect outlines the parent-child structure of the API calls.
SPNWs achieve this by extending the standard nested words model \citep{nestedwords}
with explicit fork-join structure to denote parallel blocks.
The SPNW corresponding to the execution of \cref{fig:deletetree} is:
\begin{align*}
\APIcall{delete}\ \APIcall{log}\ \APIret{log}\ \fjword{3}{w_1, w_2, w_3}\  \APIret{delete}
\end{align*}
In the above, the first two sub-executions  are identical: 
$w_1 = w_2 = \APIcall{review}\ \APIcall{approve}\ \APIret{approve}\ \APIret{review}$,
while the third one is different: $w_3 = \APIcall{review}\  \APIret{review}$.

The above SPNW can intuitively be understood as a structured parsing of the execution. 
We start with the request to the root, recorded as $\APIcall{delete}$ and $\APIret{delete}$ enclosing the SPNW encoding of its execution. The execution of \APIname{delete}  begins with a synchronous call to leaf node $\APIname{log}$, recorded by the first matched call-return pair $\APIcall{log}~\APIret{log}$ in the inner SPNW. The execution of the  three concurrent branches to \APIname{review} 
are recorded as part of the sub-execution $\fjword{3}{w_1, w_2, w_3}$, 
where $w_i$ denotes the SPNW corresponding to each concurrent branch. 
Notice that the first and second branches nest the execution of \APIname{approve} within that of its parent \APIname{review}. To summarize, the matched call-return pairs record the nesting in the tree, the concatenation records the sequential execution, and the fork-join block records the concurrency structure of the tree.   

This view avoids committing to a specific interleaving. It is the right view for distributed, incremental monitoring where the monitor tracks the call-return and fork-join structure as the events occur, rather than assuming a global observer that can give the entire tree or a linearized view of the execution. 

\subsection{SafePar specification}
We now illustrate series-parallel nested word expressions (SPNWEs) that we use
to express SafePar policies. 
Consider again the safe deletion policy we outlined above.
Intuitively, it can be decomposed into two phases:  
(a) the logging phase, and (b) the concurrent review phase.

\newpar{Logging phase}
The accountability requirement asks that a delete request be logged. 
We can encode this requirement as a mini-policy consisting of the matched call-return pair for \APIname{log}: $p_{\APIname{log}}~=~\APIcall{log}~\APIret{log}$.

\newpar{Single approved review}
A review approval requires \APIname{review} to synchronously call \APIname{approve} as its child. The execution of \APIname{approve} is nested between that of \APIname{review}. We express this parent-child relationship by \textit{nesting} a well-matched call to \APIname{approve}, \textit{i.e.,} $\APIcall{approve}~\APIret{approve}$ between the call and return of its parent \APIname{review} API: $p_{\APIname{approve}} = \APIcall{review}\ \APIcall{approve}\ \APIret{approve}\ \APIret{review}$.

\newpar{Concurrent review phase}
The review phase consists of several concurrently executing \APIname{review} branches. The policy requires that at least two of these branches should satisfy $p_{\APIname{approve}}$. In SafePar, such collective constraints across parallel branches are expressed using  a parallel composition operator $\parnew(\ldots)$ that takes a constraint, like $p_{\APIname{approve}}$, with a multiplicity guard, like $\gemult{2}$: 
$p_{\APIname{review}} ~=~\parnew(p_{\APIname{approve}}^{\gemult{2}})$. The guard
requires that at least two branches in the parallel block must match $p_{\APIname{approve}}$. Other branches in the same block may fail to match this branch constraint without violating the safety policy.

\newpar{Composing the logging and the review phases}
The deletion policy first requires that logging must complete before the
concurrent review phase. We express this condition by sequentially composing the sub-policies:
$
p_{\APIname{log}}~p_{\APIname{review}}$.
Next, since the entire workflow is part of \APIname{delete}'s execution, we express the full policy as:  
$p_{\textsf{safeDel}} = \APIcall{delete}~p_{\APIname{log}}~p_{\APIname{review}}~\APIret{delete}$.

This policy illustrates three aspects of our language: (a) call-return nesting, (b) sequential composition, and (c) parallel operator to express constraints over the concurrent branches. 

\subsection{SafePar policy enforcement}
SafePar enforces an SPNWE policy by compiling it to \textit{series-parallel visibly pushdown automaton} (SP-VPA)
that we introduce in this work. 
SP-VPA is an extension of standard VPA to support aggregate summaries from
parallel blocks. It preserves the usual visible stack discipline of a VPA for
synchronous API calls and returns, where reading a call symbol pushes a stack
symbol and reading a return pops it.
For example, for the delete policy, in a successful run of the automaton on the above SPNW, the automaton will behave like a usual VPA until the fork-join block. In the parallel block, the automaton tracks each branch in parallel. The SP-VPA aggregates the summaries of the automaton run on individual branches and checks whether at least two branches have satisfied the review approval policy. Finally, the automaton transitions on $\APIret{delete}$ like a usual VPA.  

SafePar realizes this enforcement in the servicemesh networking
layer~\citep{ashok21servicemeshes,servicemesh}, where each service's container is co-located with a sidecar container running a proxy that intercepts all the API call and returns of the given service. The proxy observes API calls and returns, simulates the corresponding SP-VPA transition, and propagates the current monitor configuration through HTTP headers. This lets SafePar's monitor enforce policies in a blackbox and non-invasive manner. 


\section{Series-Parallel Nested Words}
\label{sec:syntax}

In this section, we formally define series-parallel nested words (SPNWs), 
a word model for encoding  microservice traces with both synchronous and asynchronous API calls. 
We then  present the specification language used by the  
SafePar framework for expressing concurrent inter-service communication properties.

As discussed in \cref{sec:overview}, a runtime monitor in the microservice setting does not receive the entire execution tree as an input object. Instead, it observes a stream of HTTP call and return events online. Therefore, we need a word-like trace representation that does not impose a global linearization on concurrent events, while still preserving the structure needed to enforce useful safety properties. These include: (a) matched API calls and returns, which define each API's execution scope; (b) parent-child nesting; (c) sequential order, which  records when one sibling execution completes before the next sibling begins executing; (d) fork-join structure, which records which sibling APIs execute in parallel. 

Nested words \cite{nestedwords} satisfy the first three requirements by enriching a linear sequence of API call and return symbols with information about the matching call/return symbol. This is sufficient for the synchronous setting, where sibling APIs sequentially execute one after another. However, nested words cannot distinguish sequential sibling calls from concurrently executing siblings. For instance, a nested word can express that, in \cref{fig:shorttree}, both the \APIname{review} APIs are invoked inside the scope of \APIname{delete};
it cannot, however, express that they execute concurrently. This distinction is important for policies that require enforcing constraints such as `enough concurrent replicas should respond', 
`a parallel read to database should also be logged', or 
`the number of parallel calls performing some sensitive operation is bounded above by $c$' for some fixed number $c$.
In fact, when the asynchronous structure of executions is not faithfully
represented, then the structure may incorrectly be labelled as ill-nested!
SPNWs, which extend nested words with an explicit parallel composition operator, allows us to 
precisely capture asynchrony and nesting.

\subsection{Series-Parallel Nested Word Grammar}
\boldpara{Notation.}
Let $\basealpha$ be the set of API names. For an API $\APIname{a} \in \basealpha$, we write $\APIcall{a}$ for its call symbol and $\APIret{a}$ for its matching return symbol. Let the set of all API call events be $\Sigma_c = \{\APIcall{a} \mid \APIname{a} \in \basealpha\}$ and the set of all API return events be $\Sigma_r = \{\APIret{a} \mid \APIname{a} \in \basealpha\}$. We write $\Sigma = \Sigma_c \cup \Sigma_r$.

\begin{definition}
The set of SPNWs over $\Sigma$ is defined by the grammar:
 \begin{align*}
 w ::= \epsilon~\vert~\APIcall{a} ~w~\APIret{a} ~\vert ~ w_1 \cdot w_2 ~ \vert ~ \fjword{n}{w_1, \ldots , w_n}, ~\textit{where}~ \APIname{a} \in \basealpha \text{ and } n  \ge 2.
 \end{align*}
\end{definition}

Here, $\epsilon$ denotes the empty trace. The word $\APIcall{a}~w~\APIret{a}$ denotes an execution trace of API $\APIname{a}$: the terminal call and return events mark the beginning and ending of the API's execution, respectively, and the subword $w$ record the events that occur during this API's  execution. Sequential composition $w_1 \cdot w_2$ denotes that the execution represented by $w_1$ completes before the execution represented by $w_2$ begins. The parallel composition $\fjword{n}{w_1, \ldots, w_n}$ denotes a parallel (fork-join) block of $n$ concurrent sibling executions. Here, the symbol $\fork_n$ opens the parallel block and records its arity; the symbol $\join$ closes the block after all the branches complete. 
Each $w_i$ represents the SPNW for one branch.
Branches are unordered and as meant to execute concurrently with each other.
Further, all the branches complete their execution before the surrounding execution continues. 
Thus, SPNWs extend nested words with an explicit parallel 
composition operator while preserving the usual well-nested call-return structure of nested words.

Let us illustrate SPNWs using three different execution traces: 
a purely sequential trace, a purely parallel trace, an a trace that combines the two.

\begin{example}[Sequential Composition]\label{nestseq}
Consider an execution in which API $\APIname{delete}$ first calls one $\APIname{review}$; and after this review returns, $\APIname{delete}$ calls a second $\APIname{review}$; finally, after the second review returns, \APIname{delete} sends its response. The corresponding SPNW is:
\[
\APIcall{delete} ~\APIcall{review}~ \APIret{review}~ \APIcall{review}~ \APIret{review}~ \APIret{delete}.
\]

The events from both $\APIname{review}$ executions are nested  between the call/return symbol of the parent $\APIname{delete}$. The sequential concatenation denotes that the execution of the first $\APIname{review}$ was completed before the execution of the second $\APIname{review}$ began. 
\end{example}

The next example illustrates how SPNWs model concurrent sibling executions.

\begin{example}[Parallel Composition]\label{nestpar}
Consider an asynchronous variant of the above example, where API $\APIname{delete}$ calls both the $\APIname{review}$ APIs in parallel and returns only after both the children have returned. This trace can be represented using the parallel composition operator:
\[
\APIcall{delete} ~\fjword{2}{\APIcall{review}~ \APIret{review}, ~\APIcall{review}~ \APIret{review}} \APIret{delete}.
\]

Eventhough, the tuple lists the two sibling branches in order, as such, the fork-join combinator does not impose an order between the individual events in these two siblings.
\end{example}

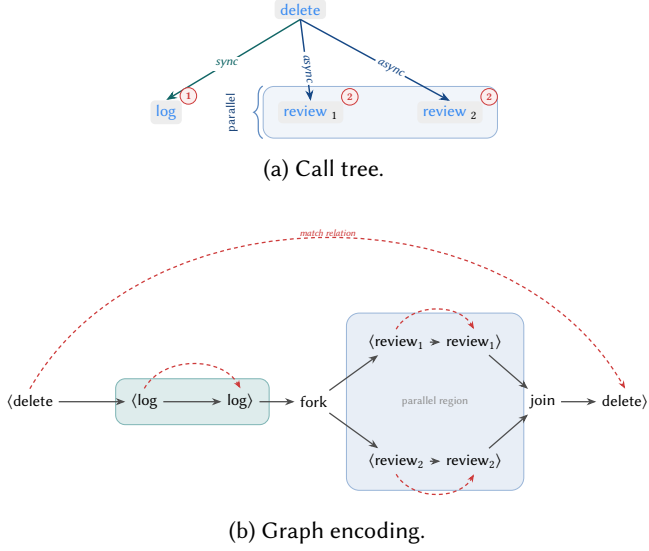
\begin{figure}[t]
\centering

\begin{subfigure}[b]{0.33\linewidth}
\centering
\snfit{\begin{tikzpicture}[
  >=Stealth,
  syncarr/.style={-{Stealth[length=6pt]}, thick, namecol!85!black},
  asyncarr/.style={-{Stealth[length=6pt]}, thick, pwblue!80!black},
  elbl/.style={font=\scriptsize\sffamily\itshape, fill=white, inner sep=1.4pt},
  stepnum/.style={circle, draw=badred, fill=badred!8, text=badred, inner sep=0pt,
                  minimum size=3.2mm, font=\tiny\bfseries},
]

\node[pev]                (delete)  at (0, 0)    {\APIname{delete}};
\node[pev]                (consent) at (-2.6,-2) {\APIname{log}};
\node[pev]                (review1) at (0.2,-2)  {\APIname{review}$_1$};
\node[pev]                (review2) at (2.9,-2)  {\APIname{review}$_2$};

\begin{scope}[on background layer]
  \node[fit=(review1)(review2), draw=pwblue!45, rounded corners=6pt,
        fill=pwblue!6, inner sep=8pt] (parblock2) {};
\end{scope}

\draw[syncarr]  (delete.south) -- node[elbl, text=namecol!85!black, pos=.55] {\textsf{sync}} (consent.north);
\draw[asyncarr] (delete.south) -- node[elbl, text=pwblue!80!black, pos=.6, sloped] {\textsf{async}} (review1.north);
\draw[asyncarr] (delete.south) -- node[elbl, text=pwblue!80!black, pos=.6, sloped] {\textsf{async}} (review2.north);

\draw[pwblue!70, line width=.7pt, decorate,
      decoration={brace, amplitude=5pt, mirror}]
  (parblock2.north west) -- (parblock2.south west)
  node[midway, xshift=-6mm, font=\scriptsize\sffamily, text=pwblue!70!black,
       rotate=90] {parallel};

\node[stepnum] at ($(consent.north east)+(1.2mm,0.8mm)$) {1};
\node[stepnum] at ($(review1.north east)+(1.2mm,0.8mm)$) {2};
\node[stepnum] at ($(review2.north east)+(1.2mm,0.8mm)$) {2};

\end{tikzpicture}}
\caption{Call tree.}
\label{fig:shorttree}
\end{subfigure}
\hfill

\begin{subfigure}[b]{0.62\linewidth}
\centering
\snfit{\begin{tikzpicture}[
  >=Stealth,
  every node/.style={font=\sffamily},
  tok/.style={font=\sffamily},
  arr/.style={-{Stealth[length=6pt]}, thick, black!70},
  matcharr/.style={-{Stealth[length=5pt]}, thick, badred, dash pattern=on 2.4pt off 2pt},
]

\node[tok] (Ropen)  at (0,0)     {$\langle$\textsf{delete}};
\node[tok] (Copen)  at (2.5,0)   {$\langle$\textsf{log}};
\node[tok] (Cclose) at (4.6,0)   {\textsf{log}$\rangle$};
\node[tok] (fork)   at (6.2,0)   {\textsf{fork}};
\node[tok] (join)   at (11.2,0)  {\textsf{join}};
\node[tok] (Rclose) at (13.0,0)  {\textsf{delete}$\rangle$};

\node[tok] (Dopen)  at (8.0,1.3)  {$\langle$\textsf{review}$_1$};
\node[tok] (Dclose) at (9.7,1.3)  {\textsf{review}$_1\rangle$};
\node[tok] (Lopen)  at (8.0,-1.3)  {$\langle$\textsf{review}$_2$};
\node[tok] (Lclose) at (9.7,-1.3)  {\textsf{review}$_2\rangle$};

\begin{scope}[on background layer]
  \node[fit=(Copen)(Cclose), fill=namecol!14, draw=namecol!55,
        rounded corners=6pt, inner sep=6pt] (seqbox) {};
  \node[fit=(Dopen)(Dclose)(Lopen)(Lclose), fill=pwblue!10, draw=pwblue!45,
        rounded corners=8pt, inner sep=10pt] (parbox) {};
\end{scope}

\draw[arr] (Ropen)  -- (Copen);
\draw[arr] (Copen)  -- (Cclose);
\draw[arr] (Cclose) -- (fork);
\draw[arr] (fork)   -- (Dopen);
\draw[arr] (Dopen)  -- (Dclose);
\draw[arr] (fork)   -- (Lopen);
\draw[arr] (Lopen)  -- (Lclose);
\draw[arr] (Dclose) -- (join);
\draw[arr] (Lclose) -- (join);
\draw[arr] (join)   -- (Rclose);

\draw[matcharr] (Ropen.north) to[out=60, in=120]
  node[midway, above, font=\tiny\itshape\sffamily, text=badred, inner sep=1pt] {match relation}
  (Rclose.north);
\draw[matcharr] (Copen.north)  to[out=60,  in=120] (Cclose.north);
\draw[matcharr] (Dopen.north)  to[out=60,  in=120] (Dclose.north);
\draw[matcharr] (Lopen.south)  to[out=-60, in=-120] (Lclose.south);

\node[font=\scriptsize\sffamily, text=black!45] at (parbox.center) {parallel region};

\end{tikzpicture}}
\caption{Graph encoding.}
\label{fig:deletegraph}
\end{subfigure}

\caption{(a) Simplified variant of the API call tree for the admin console deletion example in \cref{fig:deletetree}. \APIname{log} API  is invoked synchronously and it returns before the two asynchronous calls to \APIname{review}; (b) series-parallel graph encoding of the tree with APIs called in parallel are wrapped between \fork and \join nodes and API call/return matching depictned in red.}
\label{fig:deleteboth}
\end{figure}

Finally, we illustrate how sequential and parallel compositions can be combined.
\begin{example}[Sequence+Parallel Combination]\label{nestseqpar}
Suppose the API $\APIname{delete}$ first calls $\APIname{log}$, and when $\APIname{log}$ returns, $\APIname{delete}$ calls two $\APIname{review}$ in parallel. Finally, after both the $\APIname{review}$ calls return, $\APIname{delete}$ sends its response. This trace is encoded as follows:
\[\APIcall{delete} ~\APIcall{log}~ \APIret{log}~ \fjword{2}{\APIcall{review}~ \APIret{review}, ~\APIcall{review}~ \APIret{review}} \APIret{delete}.\]
Here, the execution of $\APIname{log}$ is sequentially composed with the parallel block containing the concurrent executions of two $\APIname{review}$.
\end{example}
This SPNW encodes the execution trace in \cref{fig:shorttree}. To highlight the series-parallel structure of the word, \cref{fig:deletegraph} visualizes  the same trace as a series-parallel graph. Reading from left to right, the graph begins with $\APIcall{delete}$ and ends with its matching return, shown by the red dashed arc. Nested between these terminal,  $\APIname{delete}$ first invokes $\APIname{log}$ synchronously, represented by the matched pair $\APIcall{log}\;\APIret{log}$. The trace then reaches the \fork node, which corresponds to $\APIname{delete}$  API starting a fork-join block for two concurrent executions of $\APIname{review}$. The two outgoing branches  execute independently and synchronize at $\join$ before the control returns to $\APIname{delete}$.

\subsection{Series-Parallel Nested Word Expressions}

We now describe \textit{series-parallel nested word expressions} (SPNWEs),
which is a class of rational expressions.
Intuitively, SPNWEs correspond to \emph{regular} sets of SPNWs. 
SPNWEs extend the usual regular expression operations for choice, 
sequencing, and iteration with a parallel composition operator 
to specify constraints over the branches of a fork-join block. 

The key feature of the parallel composition operator is that it does not fix the number or order of concurrently executed branches. Instead, each operand of a parallel operator describes a class of branch executions and is annotated with  a \textit{multiplicity constraint} describing how many branches of that class must appear in the parallel block. These multiplicity constraints are necessary because the number of branches spawned in a parallel block is a dynamic property of the execution. For example, a service may fan out to an input-dependent number of replicas, log, database. Therefore, a policy may require that ``at least one branch logs the operation'', ``exactly one branch performs a payment'' or ``no branch accesses private database''.
Formally, a multiplicity constraint $m = (\bowtie, n)$ (or simply $m = \bowtie n$)
is a pair of a comparison operator $\bowtie \in \{=,\le,\ge\}$
and a threshold $n \in \mathbb{N}$. For a number $i \in \mathbb{N}$,
we write $i \vDash ~\bowtie~n$ iff $i~ \bowtie~n$  holds.

\begin{definition}
The syntax of SPNWEs is defined by the grammar: 
 \begin{align*}
 spe ::= 0~\vert~1 ~\vert~\APIcall{a}~ \cdot spe \cdot \APIret{a}~ \vert ~ spe_1 + spe_2 ~\vert ~ spe_1 \cdot spe_2 ~\vert ~ spe^* ~ \vert ~ \parnew (spe_1^{m_1}, \ldots, spe_k^{m_k}).
 \end{align*}
\end{definition}
 
The expression $0$ denotes the null expression, and corresponds to the empty set of SPNWs.
$1$ denotes the unit expression and corresponds to the language containing the empty word. 
The expression $\APIcall{a} \cdot spe \cdot \APIret{a}$ specifies that the body of  API $\APIname{a}$'s execution should satisfy $spe$. 
We write a leaf call to $\APIname{a}$ as the shorthand $\APIcall{a}\ \APIret{a}$ for $\APIcall{a}~1~\APIret{a}$. 

The expression $spe_1 + spe_2$ specifies a choice: an execution may satisfy either $spe_1$ or $spe_2$. The expression $spe_1 \cdot spe_2$ denotes sequencing: an accepted word is a concatenation of a word accepted by $spe_1$ followed by a word accepted by $spe_2$. The Kleene star $spe^*$ expression denotes finite iteration of words accepted by $spe$. 

The parallel composition expression $\parnew (spe_1^{m_1}, \ldots, spe_k^{m_k})$ matches a parallel fork-join block when its branches can be assigned  to the sub-expressions $spe_1, \cdots, spe_k$ such that, for every $i$, the number of branches assigned to $spe_i$ satisfies the multiplicity constraint $m_i$. The expression only  constrains the collection of concurrent branches, not their relative execution order.
For instance, the expression $\parnew(spe^{\ge 5})$ specifies that in a parallel block of several  concurrent branches, there should be atleast five branches satisfying $spe$. Other branches are not requires to satisfy $spe$. 

Now, we formally define the semantics of SPNWEs. Below we write $[n] \triangleq \{1,\ldots,n\}$, and for a function $\alpha: A \to B$ and an element $j \in B$, we let the pre-image of $j$ be $\alpha^{-1}(j) \triangleq \{i \in A \mid \alpha(i) = j\}$.

\begin{definition}
The language $L(spe)$ of an SPNWE spe is defined inductively as follows:
 \begin{align*}
 &L(0) \triangleq \emptyset\\
 &L(1) \triangleq \{\epsilon\}\\
 &L(\APIcall{a} \cdot spe \cdot \APIret{a}) \triangleq \{\APIcall{a} ~w ~\APIret{a} \mid w \in L(spe)\}\\
 &L(spe_1 + spe_2) \triangleq L(spe_1) \cup L(spe_2)\\
 &L(spe_1 \cdot spe_2) \triangleq \{w_1 \cdot w_2 \mid w_1 \in L(spe_1), ~w_2 \in L(spe_2)\}\\
 &L(spe^*) \triangleq \bigcup_{i \ge 0} L(spe)^i , \text{where} ~L(spe)^{i+1} = L(spe) \cdot L(spe)^{i}, ~\text{and}~ L(spe)^0 = \{\epsilon\}\\
 &L(\parnew (spe_1^{m_1}, \ldots, spe_k^{m_k})) \triangleq
 \left\{\, \fjword{n}{w_1, \ldots, w_n} ~\middle|~
 \begin{array}{@{}l@{}}
 n \in \mathbb{N},\\
 \exists \alpha: [n] \to [k] \cup \{\bot\}\ \text{s.t.}\\
 \alpha(i) = \bot \Rightarrow \forall j.~w_i \notin L(spe_j)\quad(i \in [n]),\\
 \alpha(i) \neq \bot \Rightarrow w_i \in L(spe_{\alpha(i)})\quad(i \in [n]),\\
 |\alpha^{-1}(j)| \vDash m_j\quad(j \in [k])
 \end{array}
 \,\right\}
 \end{align*}
\end{definition}

Here, the expression $\APIcall{a} \cdot spe \cdot \APIret{a}$ accepts SPNWs where the inner sub-word between the outer $\APIcall{a}$ and its matching $\APIret{a}$ symbol is accepted by $spe$. The interpretation for the choice, sequence, and Kleene star follow the usual regular expression interpretation. The parallel expression accepts a word if there exists an assignment $\alpha:[n]\to [k] \cup \{\bot\}$ from the branch $w_i$ to the sub-expression $spe_{\alpha(i)}$ that accepts the word such that the number of branches satisfying $spe_j$, \textit{i.e.,} $|\alpha^{-1}(j)|$, should satisfy the constraint $m_j$. Branches that do not match any of the listed sub-expressions are assigned to $\bot$ and are ignored by the multiplicity constraints. The acceptance condition for the parallel word requires the existence of some valid $\alpha$.

We illustrate SPNWEs using three common patterns: nesting, iteration, and parallel composition. 

\begin{example}[Nesting]
Consider API $\APIname{A}$ is allowed to invoke only one child API  $\APIname{B}$, which must immediately return without making any further calls. First, we express that API $\APIname{B}$ should not invoke any APIs as $\APIcall{B}~\APIret{B}$, a shorthand for $\APIcall{B}~1~\APIret{B}$. Then we specify the parent-child constraint as: \[\APIcall{A}~\APIcall{B}~\APIret{B} ~ \APIret{A}.\]
\end{example}

\begin{example}[Kleene Star]
Suppose in the previous example $\APIname{A}$ may invoke any number of leaf calls to $\APIname{B}$ or $\APIname{C}$, one after the other, sequentially. We specify this using the union and Kleene star operator as: \[\APIcall{A}~(\APIcall{B}~\APIret{B} + \APIcall{C}~\APIret{C})^* ~ \APIret{A}.\]
\end{example}
Next, we specify a constraint on concurrent executions.
\begin{example}[Parallel composition]
Consider that in a parallel block, exactly one branch should satisfy $spe_1$ and no branch should satisfy $spe_2$. This can be expressed as: \[\parnew(spe_1^{\eqmult{1}}, spe_2^{\eqmult{0}}).\]
Note that this policy does not impose any order in which branches should satisfy $spe_1$ and $spe_2$.
\end{example}
Finally, we describe an example combining policies over synchronously and asynchronously invoked children.
\begin{example}[Sequence and parallel combination]
 Consider that API $\APIname{A}$ should invoke some concurrent APIs such that exactly two branch should match $spe_1$ and no branch should match $spe_2$; once these concurrent APIs return, $\APIname{A}$ should invoke $\APIname{B}$, which should immediately return; finally $\APIname{A}$ also returns. Specifying this property requires sequentially combining the children constraints in the previous examples as follows:
 \[\APIcall{A}~\parnew(spe_1^{\eqmult{2}}, spe_2^{\eqmult{0}})~\APIcall{B}~\APIret{B} ~ \APIret{A}.\]
\end{example}
Let $w_1, w_2 \in L(spe_1)$ and $w_2 \in L(spe_2)$. 
The above property accepts $\APIcall{A}~\fjword{2}{w_1,w_2}~\APIcall{B}~\APIret{B} ~ \APIret{A}$ because there exists a branch assignment $\alpha \triangleq \{1\mapsto 1, 2\mapsto 1\}$ under which both  branches are matched by $spe_1$. This assignment witnesses that two branches satisfied the required multiplicity for $spe_1$, while no branch is assigned to $spe_2$. However, the second branch also matches $spe_2$, so the branch assignment is not uniquely determined by the trace. Thus, an online monitor cannot deterministically pick the correct assignment when a branch can match multiple sub-expressions. Since our goal is to deterministically monitor policies, we restrict SafePar specifications to a monitorable fragment of SPNWEs.

\section{SafePar Language: A monitorable fragment of SPNWEs} \label{sec:casestudies}
We first define the auxiliary notions used to define the SafePar fragment. Nullability records whether an expression accepts the empty word. Since the same symbol may repeat in an expression, we distinguish different syntatic occurrences of symbols by assigning each occurrence a unique position. For instance, in the SPNWE $spe = \APIcall{a}^{1} \APIcall{a}^{2} \APIret{a}^{3} \APIret{a}^{4}$, we have uniquely identified each syntactic occurrence of calls and returns to $\APIname{a}$ by the superscripts. We write $Pos(spe) = \{\APIcall{a}^{1}, \APIcall{a}^{2}, \APIret{a}^{3}, \APIret{a}^{4}\}$ for the set of positions and $\poslabel(p)$ for the symbol at the position $p$. For instance, $\poslabel(\APIcall{a}^2) = \APIcall{a}$. 

We also require three standard position sets to state 1-unambiguity: $\First$, $\Follow$, and $\Last$;
their formal definitions are presented in Appendix, \appref{sec:appendix} but
we describe them intuitively here.
For an expression $spe$, $\First(spe)$ is the set of positions that can appear as the first event of some word in $L(spe)$. For example, in the above example $\First(spe) = \{\APIcall{a}^1\}$. Similarly, $\Last(spe)$ is the set of positions that can appear as the last symbol of some word in $L(spe)$. In the above example, $\Last(spe) = \{\APIret{a}^4\}$. For a position $p \in Pos(spe)$, the set $\Follow_{spe}(p)$ has positions that can immediately follow $p$ in some word in $L(spe)$. For instance, in the above example, $\Follow_{spe}(\APIcall{a}^2) = \{\APIret{a}^3\}$.

\begin{definition}[SafePar specifications]
The \textsf{SafePar} fragment is the smallest set of SPNWEs satisfying the following conditions:
\begin{enumerate}
\item $0$, $1$ are in \textsf{SafePar}.
\item $\APIcall{a} ~spe'~\APIret{a}$ is in \textsf{SafePar} if $spe'$ is in \textsf{SafePar}.
\item $spe_1 + spe_2$ is in \textsf{SafePar} if $spe_1$ and $spe_2$ are in \textsf{SafePar} and $\poslabel(\First(spe_1))\cap \poslabel(\First(spe_2))= \emptyset$ and not both $spe_1$ and $spe_2$ are nullable.
\item $spe_1 \cdot spe_2$ is in \textsf{SafePar} if $spe_1$ and $spe_2$ are in \textsf{SafePar}  and  for every $p \in \Last(spe_1)$, $\poslabel(\Follow_{spe_1}(p))\cap \poslabel(\First(spe_2))=\emptyset$. Moreover, if $spe_1$ is nullable then $\poslabel({\First(spe_1)}) \cap \poslabel(\First(spe_2)) = \emptyset$.
\item $spe^*$ is in \textsf{SafePar} if $spe$ is non-nullable and in \textsf{SafePar}, and for every $p \in \Last(spe)$ satisfy $\poslabel({\Follow_{spe}(p)}) \cap \poslabel({\First(spe)})=\emptyset$. 
\item $\parnew (spe_1, \ldots, spe_k)$ is in \textsf{SafePar} if every $spe_i$ is non-nullable and in \textsf{SafePar} and for all distinct $i, j \in [k]$, $\poslabel({\First(spe_i)}) \cap \poslabel({\First(spe_j)})= \emptyset$.
\end{enumerate}
\end{definition}
Here, the union condition ensures that the next observed symbol uniquely determines which sub-expression matches it, while the nullability condition rules out an ambiguous empty choice. The concatenation condition ensures that, after reading a position matched by $spe_1$, the next symbol cannot be interpreted as a continuation of a word matching $spe_1$  and the starting symbol of words matching $spe_2$. The additional nullable case handles the situation where $spe_1$ may be skipped. The Kleene star condition ensures that the next symbol uniquely determines whether it is a continuation of the current iteration or starting of a new one. Finally, the parallel condition ensures that the first observable symbol of each branch uniquely determines which sub-expression can that branch match. If the first symbol of a branch is not in the set of first symbols of any sub-expression, the branch is irrelevant and assigned to $\bot$. Thus, the witness function $\alpha$ that recorded the expression a branch matched can be uniquely picked. Besides the parallel condition, these requirements closely follow the standard 1-unambiguity conditions for regular expressions~\cite{unamb}.

\subsection*{Case Studies}
We illustrate SafePar's expressiveness through representative policies that combine constraints on nesting, sequencing, and the collective behavior of concurrent branches.

\paragraph*{\textbf{Scoped order of children.}}
SafePar can express ordering constraints among the children API calls of a parent request. This is useful for: a compliance workflow that may require the data to be  encrypted before being written to a database; an audit workflow that may require logging before the parent returns; or a deployment pipeline requiring that a build artifact's Docker image should be built and scanned before being published. We illustrate this pattern with a hospital compliance policy.

\begin{csexample}
Consider a hospital application comprising: a \APIname{test} API for requesting a medical test; a  \APIname{pay} API for charging the patient; a \APIname{de-id} API for removing a patient's personal health information (PHI) from its input;  and a \APIname{lab} API for submitting the request to a third-party lab. Suppose the compliance team requires the patient to be charged first, the PHI to be de-identified next, and the request to be submitted to the third party \APIname{lab} only after de-identification. 
Let $p_{\APIname{pay}} = \APIcall{pay}\ \APIret{pay}$ denote a call to the payment API; $p_{\APIname{de-id}} = \APIcall{de-id}\ \APIret{de-id}$ denote a call to de-identify the record; $p_{\APIname{lab}} = \APIcall{lab}\ \APIret{lab}$ denote a call to the \APIname{lab} API. This property can be specified by describing the constraint on the SPNW between the API call/return of the \APIname{test} API:

\[\APIcall{test} ~ p_{\APIname{pay}}~ p_{\APIname{de-id}} ~p_{\APIname{lab}}~\APIret{test}.\]

\end{csexample}

\paragraph*{\textbf{Iterated sibling constraint.}}
We can use the Kleene star to specify that a parent may invoke multiple child API calls, each satisfying the same sub-policy. This is useful when a parent first performs a one-time setup action like authentication or secure channel establishment and then repeatedly invoke similar child operations like  multiple database writes or secure message transmissions. One such example is described below.

\begin{csexample}
Consider a file download application  comprising: a \APIname{download} API that starts a batch download request; \APIname{authenticate} API that validates the user identity; \APIname{db} API that reads files from the backend. A developer may want to enforce that $\APIname{download}$ authenticates the user before reading any file, after which the user can read any number of files during the session using the \APIname{download} API. This property can be specified using the Kleene star around the sub-property $p_{\APIname{db}} = \APIcall{db}~ \APIret{db}$ describing the call to \APIname{db}:
\[\APIcall{download} ~ \APIcall{authenticate}~ \APIret{authenticate} (p_{\APIname{db}})^*~\APIret{download}.\]
\end{csexample}
\paragraph*{\textbf{Single constraint across concurrent branches.}}
SafePar can constrain the number of concurrent child branches that satisfy a given sub-policy $p$ using a pattern of the form $\parnew(p^{m})$. Different multiplicities $m$ capture different classes of branch-level constraints.

Upper-bound multiplicities, expressed using the \textit{at most} construct, are useful for specifying resource limits. For example, a policy may bound the number of writes or notifications to avoid overwhelming the application; or limit the number of calls to a third-party service to ensure the application stays within budget. This is illustrated below. 

\begin{csexample}
Consider a storage service comprising: the frontend \APIname{save} API for saving data and \APIname{write} API for writing data to a backend database. For reliability, $\APIname{save}$ may send multiple  \APIname{write} calls in parallel to update multiple database replicas. Although this parallelism can reduce latency, too many writes from a single parent request can overload the backend. So a deployment team may enforce an upper bound on the number of \APIname{write} branches that any \APIname{save} request can invoke. Let $p_{\APIname{write}} = \APIcall{write}~\APIret{write}$ denote a successful write branch. Then the bounded write policy is specified as:
\[
\APIcall{save}~\parnew(p_{\APIname{write}}^{\lemult{k}})~\APIret{save}.
\]
\end{csexample}

SafePar's lower-bound multiplicities specify requirements such as quorum or Byzantine approval, where sufficiently many concurrent branches must succeed before the parent request continues. This is illustrated below.

\begin{csexample}
Consider a cloud admin console that executes privileged operations, such as reading confidential customer data or  deleting a production database. 
Let this functionality be implemented using: the frontend $\APIname{console}$ API that takes in the operation to be executed; the $\APIname{auth}$ API that checks if the user has admin privilege to execute a safety-critical operation; the $\APIname{run}$ API that invokes the requested operation.  Before invoking the operation, the console checks if the requesting user has the required admin privilege. 
For availability, the authorization state, such as access control lists (ACLs), may be replicated across multiple backend databases. After a user's admin privilege is revoked, some replicas may temporarily have stale ACL entries. As a result, querying a single stale replica may incorrectly approve a privileged operation.

A security team may, therefore, enforce a \textit{quorum policy}: $\APIname{console}$ must query multiple $\APIname{auth}$ replicas in parallel and call $\APIname{run}$ only after atleast $n$ replicas approve. Now, let the policy for a successful approval be  $p_{\APIname{auth}} = \APIcall{auth}~ \APIret{approved-auth}$, where the matching approval response for the authenticated call is denoted by $\APIret{approved-auth}$. Then the quorum authorization can be specified using the \textit{atleast} $n$ multiplicity on the $\APIname{auth}$ branches:
\[\APIcall{console}~ \parnew (p_{\APIname{auth}}^{\gemult{n}}) ~\APIcall{run}~ \APIret{run}~\APIret{console}, \text{where} p_{\APIname{auth}} = \APIcall{auth}~ \APIret{approved-auth}.\]
\end{csexample}

SafePar's exact constraints let us express uniqueness requirements. For example, they can require that non-idempotent operations, like payment, password update, or one-time token use, occurs exactly once within a parent request. The special case of an exact bound set to zero can be used to specify that some operation is forbidden. For instance, no branch should access private details, no file read branch should return without logging. This is illustrated below.
\begin{csexample}
Consider a password reset service implemented using: frontend $\APIname{recover}$ API to reset the password; $\APIname{email}$ API to send a link to the registered email; $\APIname{phone}$ to send a link to the registered  phone number; and $\APIname{reset}$ to update the password after clicking the link. To improve availability, the $\APIname{recover}$ service may send the recovery links through multiple registered channels, like email and SMS to all phones in parallel. Each channel can lead to a \APIname{reset} call to update the password. Since a password update is a non-idempotent security-critical action, a recovery request should contain exactly one successful password update. Let
$p_{reset} = \APIcall{email} ~\APIcall{reset}~\APIret{reset}~\APIret{email} ~+~ \APIcall{phone} ~\APIcall{reset}~\APIret{reset}~\APIret{phone}$ be the policy of a successful recovery branch capturing that an email link or a phone message was used to reset the password. Then the password recovery policy is expressed as:
\[\APIcall{recover} ~\parnew(p_{reset}^{\eqmult{1}})~\APIret{recover}.\]
The multiplicity $\eqmult{1}$ specifies that only one reset branch can exist in a recovery request. The policy rejects a trace where the password is not updated or where it is updated through multiple channels.
\end{csexample}

\paragraph*{\textbf{Multiple constraints across concurrent branches.}}
Recall that the parallel construct can express policies in which different concurrent branches must satisfy different sub-policies with different multiplicity constraints. This can be specified using a pattern of the form $\parnew (p_1^{c_1}, p_2^{c_2})$. In addition to specifying that some set of calls may occur in parallel, this pattern  also specifies how many branches should satisfy a given sub-policy. This pattern can capture coordination requirements like ``before operation \textsf{A}, exactly (or atleast)  $m_i$ branches should satisfy the behavior encoded by the policy $p_i$. Policies requiring such constraints arise in distributed workflows where progress to a later stage depends on completion of multiple concurrent operations exposed by different services.  For example, a CI/CD pipeline must provision compute and storage resources, and enable networking in parallel before it begins running the workflow; an e-commerce application should require payment authorization and inventory reservation before placing an order.

\begin{csexample}
Consider an online store that requires the $\APIname{pay}$ service to charge the customer and the $\APIname{inventory}$ service to reserve the item in the cart. Since these services manage their persistent data independently, we should not allow one service to make a durable change when the other has failed. This could  lead to charging the customer without the item being reserved or the item being reserved even when the payment has failed.   The store therefore uses a two-phase commit before completing an order. The $\APIname{checkout}$ service first asks the $\APIname{pay}$ service to prepare by authorizing the charge and the $\APIname{inventory}$ service to prepare by placing a hold on the item. Only after both services have prepared should $\APIname{checkout}$ commit the order by calling $\APIname{order}$. This policy can be specified as:  

\[\APIcall{checkout}~\parnew(p_{\APIname{pay}}^{\eqmult{1}},~ p_{\APIname{inventory}}^{\eqmult{1}})~\APIcall{order}~\APIret{order}~\APIret{checkout}, \text{where}\]

$p_{\APIname{pay}} = \APIcall{pay}~\APIret{pay}$ and $p_{\APIname{inventory}} = \APIcall{inventory}~\APIret{inventory}$.
\end{csexample}
\paragraph*{\textbf{Constraints on nested parallel branches.}}
Some applications exhibit concurrency at more than one level. A top-level request may spawn several concurrent tasks and each task may further spawn concurrent sub-tasks. The nested word structure of our policies lets us express such \textit{hierarchical concurrency}  to describe the internal structure of a concurrent branch. Such patterns are common in hierarchical coordination workflows, where a coordinator decomposes the task into smaller units of work to be executed in parallel and each sub-task must independently collect enough supporting evidence from its children branches. For example, a hiring platform may run multiple interview rounds for a candidate and each round should collect evaluations from multiple interviewers; a hospital application may process several medical cases in parallel and require opinions from multiple doctors for each case; and an incident response system may launch several investigations and each investigation may require confirmation about health status from multiple services. The next example illustrates this pattern. 
 
 \begin{csexample}
 Consider a hiring platform implemented using the following APIs: $\APIname{hire}$ API to initiate the hiring; $\APIname{round}$ API to initiate an interview round; and $\APIname{eval}$ API to send an evaluation request to an interviewer. The platform evaluates a candidate by running several interview rounds in parallel. This concurrency must be controlled at two levels. At the top level the platform should not overschedule  the candidate, so the  platform may run at most $k$ rounds. At the inner level, each round must collect evaluations from at least two interviewers in parallel before the round completes to avoid single-interviewer bias. This policy can be expressed by nesting constraints on parallel blocks. Let $p_{\APIname{eval}} = \APIcall{eval}~\APIret{eval}$  denote one interviewer evaluation. Let $p_{\APIname{round}} = \APIcall{round} \parnew(p_{\APIname{eval}}^{\gemult{2}})\APIret{round}$ denote an interview round that collects atleast two evaluations. We can write the overall policy limiting the number of rounds to $k$ at the top level as:
 \[\APIcall{hire} ~\parnew(p_{\APIname{round}}^{\lemult{k}})~\APIret{hire}.\]
 
This policy rejects a trace with more than $k$ rounds. It also rejects a trace with fewer than two interviewer evaluations in any round.
 \end{csexample}
\paragraph*{\textbf{Multiple options for constraints.}}
SPNWEs can express choice between several allowed execution patterns using the union operator. This is useful when a service may support multiple mutually exclusive protocols for completing the same logical operations. Some example domains where the need for such patterns arise include authentication systems with multiple login mechanisms, storage systems that may service a request either from a cache or from a backend store.

\begin{csexample}
Consider an authentication service implemented using: $\APIname{auth}$ API that starts an authentication request, \APIname{pwd} API that offers password-based login flow, and \APIname{auth} API that supports an authenticator-based login flow. Suppose a valid authentication request should follow either the password-based login flow or the authenticator-based login flow, but it should not mix both protocols within one request. 
Suppose $p_1 = \APIcall{pwd}\ \APIret{pwd}$ describes a valid password-based login and $p_2=\APIcall{auth}\ \APIret{auth}$ describes a valid authenticator login. Then the overall policy can be specified as $\APIcall{auth} ~(p_1 + p_2)~\APIret{auth}$.
\end{csexample}

\paragraph*{\textbf{Sequential composition of series-parallel blocks.}}
We can sequentially compose constraints on series/parallel blocks with that of another to specify workflows with ordered phases. This is useful to describe settings where an application may exploit concurrency within each phase, but one phase should complete before the next starts. Such requirements arise in specifying correctness of map-reduce jobs, plan-then-execute agent workflows, cloud deployment pipelines, etc. The example below illustrates this pattern.
\begin{csexample}
Consider an agentic system implemented using the following APIs: \APIname{agent} API starts an agent run, \APIname{plan} API generates a plan for a prompt, \APIname{execute} API executes a step in the plan, \APIname{safe} checks whether an atomic action in a step is safe, and \APIname{run} API runs the action.
The $\APIname{agent}$ first runs a planning phase in which it may invoke atmost $k$ $\APIname{plan}$ API calls in parallel. Only after the planning phase completes, the agent can start the execution phase where it may issue upto $l$ $\APIname{execute}$ calls in parallel. Furthermore, each execution branch must check the safety  of each action using $\APIname{safe}$ before  running it using $\APIname{run}$.
First, we can specify the execution phase as $p_{safe} = \APIcall{execute}~\APIcall{safe}~\APIret{safe}~(\APIcall{run} ~\APIret{run})^*)~\APIret{execute}$. This requires every \APIname{run} call in an execution phase to be preceded by a call to \APIname{safe}. Second, the planning phase can be specified as $p_{plan} = \APIcall{plan}\APIret{plan}$. Finally, the sequential ordering on the planning and execution phase in agent's run can be specified as:
\[\APIcall{agent}~\parnew(p_{plan}^{\lemult{k}})~\parnew(p_{safe}^{\lemult{l}})~\APIret{agent}.\]
This policy will prevent an unsafe action from being  executed. 
\end{csexample}


\section{Visibly Pushdown Automaton for Series-Parallel Nested Words}
\label{sec:semantics}
To recognize series-parallel nested words, we define a series-parallel visibly
pushdown automaton (SP-VPA), extending the stack-based discpline of standard
visibly pushdown automaton \cite{vpa} from nested words to SPNWs. Our automata
handle sequential call/return structure exactly as in a VPA: call symbols push
stack symbols and return symbols pop stack symbols. The new aspect in SP-VPA is
its support for parallel composition: when an SP-VPA reads the parallel fragment
$\fjword{n}{w_1, \ldots, w_n}$ in an SPNW, the automaton forks into $n$
independent sub-runs, one per branch. At the synchronization point, the
automaton applies a new \textit{join transition} to the multiset of the terminal
states of each sub-run. 

We are now ready to formalize SP-VPAs. These automata have transitions specified
by four functions: $\delta_c$ for call symbols, $\delta_r$ for return symbols,
$\delta_f$ for fork symbols, and $\delta_j$ for join symbol.

\begin{definition}[Series-Parallel Visibly Pushdown Automaton]
 A \emph{series-parallel visibly pushdown automaton} (SP-VPA) $\mathcal{A}$ is a tuple $(Q, q_{init}, F, \Sigma, \Gamma, \bot, \delta_c, \delta_r, \delta_f, \delta_j, ~\kappa)$, where:
\begin{itemize}
 \item $Q$ is the set of all states, $q_{init} \in Q$ is the initial state, and $F \subseteq Q$ is the set of final states,
 \item $\Sigma = \Sigma_c  \uplus \Sigma_r$ is the alphabet, a disjoint union of call and return annotated symbols from some base alphabet $\basealpha$, 
 \item $\Gamma$ is the set of stack symbols, with a special bottom of stack symbol $\bot \in \Gamma$,
 \item $\delta_c: \calltype$ is the \textit{call transition} function,
 \item $\delta_r: \rettype$ is the \textit{return transition} function,
\item $\delta_f: Q \to Q$ is the \textit{fork} function,
\item $\delta_j: Q \times \mathcal{M}_{\kappa}(Q) \to Q $ is the \textit{join}
  function, where $\mathcal{M}_{\kappa}(Q)$ is the set of multisets over $Q$
  with all multiplicities at most $\kappa \in \mathbb{N}$.
\end{itemize}
We will exclusively consider \emph{finite} SP-VPA, where all sets are assumed to
be finite. For convenience, we will also assume that $Q \subseteq \Gamma$
throughout.
\end{definition}

The call and return transition functions, $\delta_c$ and $\delta_r$, are
inherited from VPA. On a call symbol, $\delta_c$  takes the current state and
returns the next state along with a (non-bottom) stack symbol to  push. On a
return symbol, $\delta_r$ takes the current state, the return symbol, and the
(non-bottom) symbol on top of the stack, and returns the next state. 

The fork transition function $\delta_f$ and the join transition function
$\delta_j$ are new to SP-VPA.  On a $\fork_n$ symbol, $\delta_f$ takes the
current state and returns the initial state that will be used for each of the
$n$ forked branches. On a \join symbol, $\delta_j$ takes a state and a multiset
of states---intuitively, the terminal states of branches at
a \join symbol---and returns the next state. Since the set
$\mathcal{M}_{\kappa}(Q)$ is finite, the join transition function is finitely
representable.
\begin{example}
The SP-VPA shown in \cref{fig:spvpa} starts at $q_s$ and ends at $q_A$. A solid
transition from $q_s$ to $q_0$ labeled with $\APIcall{A}/\gamma_{A}$ depicts a
$\delta_c$ transition from $q_s$ to $q_0$ on the call symbol $\APIcall{A}$ that
pushes  $\gamma_{A}$ on the stack. A solid transition from $q_{BC}$ to $q_{A}$
labeled with $\APIret{A},~\gamma_{A}$ depicts a $\delta_r$ transition from
$q_{BC}$ to $q_{A}$ on the return symbol $\APIret{A}$ when $\gamma_{A}$ is at
the top of the stack. The dashed transition from $q_{0}$ to $q_{f}$ labeled with
$\fork$ is a $\delta_f$ transition from $q_0$ to $q_f$. The two dashed
transitions labeled \join correspond to $\delta_j$ transitions. In this case,
the transition is from the state $q_0$ before the \fork to $q_{\APIname{BC}}$,
given the multiset with one occurrence of each $q_{\APIname{B}}$ and
$q_{\APIname{C}}$.
\end{example}

\begin{figure}
\centering
\begin{tikzpicture}[font=\scriptsize,
    baseline=1ex, shorten >=.4pt, node distance=14mm, on grid,
    semithick, auto,
    every state/.style={fill=white, draw=black, circular drop shadow,
        inner sep=.15mm, text=black, minimum size=7mm},
    accepting/.style={fill=gray,text=white}]

  \node (qs)  [state, initial]             {$q_s$};
  \node (q0)  [state, right=of qs]         {$q_0$};
  \node (qf)  [state, right=of q0]         {$q_f$};
  \node (q11) [state, above right=of qf]   {$q_1$};
  \node (q21) [state, below right=of qf]   {$q_2$};
  \node (q12) [state, right=of q11]        {$q_B$};
  \node (q22) [state, right=of q21]        {$q_C$};
  \node (q3)  [state, below right=of q12]  {$q_{BC}$};
  \node (q4)  [state, accepting, right=of q3] {$q_A$};

  \path [-stealth, thick]
    (qs)  edge [] node [above] {$\langle \textsf{A}/\gamma_A$} (q0)
    (q0)  edge [dashed] node [above] {\fork} (qf)
    (qf)  edge [] node [above left] {$\langle \textsf{B}/\gamma_B$} (q11)
    (qf)  edge [] node [below left] {$\langle \textsf{C}/\gamma_C$} (q21)
    (q11) edge [] node [above] {$\textsf{B}\rangle,\ \gamma_B$} (q12)
    (q21) edge [] node [below] {$\textsf{C}\rangle,\ \gamma_C$} (q22)
    (q12) edge [dashed] node {\join} (q3)
    (q22) edge [dashed] node {\join} (q3)
    (q3)  edge [] node [above] {$\textsf{A}\rangle,\ \gamma_A$} (q4)
  ;

\node [right=9mm of q4, align=left, anchor=west] {$
    \begin{aligned}
      \delta_c &= \{(q_s, \langle \textsf{A}) \mapsto (q_0, \gamma_A),\\
               &\phantom{{}=\{} (q_f, \langle \textsf{B}) \mapsto (q_1, \gamma_B),\\
               &\phantom{{}=\{} (q_f, \langle \textsf{C}) \mapsto (q_2, \gamma_C)\}\\[4pt]
      \delta_r &= \{(q_1, \textsf{B}\rangle, \gamma_B) \mapsto q_B,\\
               &\phantom{{}=\{} (q_2, \textsf{C}\rangle, \gamma_C) \mapsto q_C,\\
               &\phantom{{}=\{} (q_{BC}, \textsf{A}\rangle, \gamma_A) \mapsto q_A\}\\[4pt]
      \delta_f &= \{q_0 \mapsto q_f\}\\[4pt]
      \delta_j &= \{(q_0, \{q_{B}^1, ~q_{C}^1\}) \mapsto q_{BC}\}
    \end{aligned}$
  };
\end{tikzpicture}
\caption{A SP-VPA that accepts $L(\APIcall{A} ~\parnew((\APIcall{B}~\APIret{B})^{\eqmult{1}}, ~(\APIcall{C}~\APIret{C})^{\eqmult{1}})~\APIret{A})$.}
\label{fig:spvpa}
\end{figure}

\subsection{SP-VPA transition semantics}
We design the SP-VPA semantics to enable an \emph{incremental} online monitor,
rather than for an offline acceptor that receives the whole SPNW at once. At
each step, the monitor observes one event from each active branch. In order to
track which event occurs in which branch, we first formalize structured inputs to
the automaton that records which events belongs to which branch. Then, we define
SP-VPA transition semantics over these inputs.

We begin with the inputs. In the sequential
portions of an execution with one active branch, the automaton observes the next
call or return symbol. Inside a parallel block, the automaton observes one token
for each active branch. We formalize these structured inputs to the automaton as
the set $\token$ of \textit{incremental tokens} generated by the following grammar:
\begin{align*}
&s \in \Sigma \cup \{\fork_n \mid n \in \mathbb{N}\} \cup \{\join, \epsilon\}\\
&e ::= s \mid \parnew(e_1, \ldots, e_n)
\end{align*} 
A flat sequential token $s$ can be a single call, return, fork, join, or an empty observation $\epsilon$. A compound token $\parnew(e_1, \ldots, e_n)$ records that among the $n$ parallel branches in a parallel block, the $i^{th}$ branch observes the incremental token $e_i$. 

We will define an operational semantics for SP-VPA, where the current state of
the SP-VPA is captured by a \emph{configuration}, and each incremental token
steps the current configuration.  Outside a parallel regions, an SP-VPA
configuration is the usual VPA configuration: a state and stack. Inside a
parallel region, an SP-VPA configuration tracks the multiset of local
configurations of each active branch along with the suspended outer context that
will be resumed at the matching join.

\begin{definition}
SP-VPA configurations are generated by the following grammar:
 \begin{align*}
 T ::= \langle q, \sigma \rangle ~\vert~\langle \parnew (T_1, \ldots, T_n), \sigma \rangle
 \end{align*}
 We write $Config$ for the set of all configurations. 
\end{definition}
 
An sequential configuration $\langle q, \sigma \rangle$ consists of a state $q$
and a stack $\sigma \in \bot \Gamma^*$. A parallel configuration $\langle
\parnew (T_1, \ldots, T_n), \sigma \rangle$ consists of one local configuration
per active branch and an outer stack $\sigma$ that records the suspended parent
context to resume at the matching join; the top symbol in $\sigma$ records the
fork state $q$, while the rest of the $\sigma$ records the stack when the fork
occurred.

Now, we are ready to define the SP-VPA transition semantics.
 \begin{definition}[Transition Semantics]
 For each $e \in E$, the relation $\sptrans{e}{A} \subseteq Config \times Config$ relates the current configuration to the next configuration after reading an incremental token $e$.
 \begin{mathpar}
  \inferrule*[right=T-Call]{(q', s) \in \delta_c(q, \APIcall{a})}{\spconfig{q}{\sigma} \sptrans{\APIcall{a}}{A} \spconfig{q'}{\sigma s}}
  \and
  \inferrule*[right=T-Ret]{q' \in \delta_r(q, \APIret{a}, s)}{\spconfig{q}{\sigma s} \sptrans{\APIret{a}}{A} \spconfig{q'}{\sigma}}
  \and
  \inferrule*[right=T-Eps]{ }{\spconfig{q}{\sigma} \sptrans{\epsilon}{A} \spconfig{q}{\sigma}}
  \and

  \inferrule*[right=T-Fork]{ \delta_f(q) = q'}{\spconfig{q}{\sigma} \sptrans{\fork_{n}}{A} \spconfig{\parconfig{\underbrace{\spconfig{q'}{\bot}, \ldots , \spconfig{q'}{\bot}}_{n}}}{\sigma q}}
  \and
  \inferrule*[right=T-Par]
    {\text{for all} ~i \in [n],~T_i \sptrans{e_i}{A} T_i'}
    {\spconfig{\parconfig{T_1, \ldots , T_n}}{\sigma} \sptrans{\parword{e_1, \ldots , e_n}}{A} \spconfig{\parconfig{T_1', \ldots , T_n'}}{\sigma}}
   \and
   \inferrule*[right=T-Join]{M = \trunc{k}{\multiset{q_1, \ldots , q_n}} \quad q' \in \delta_j(q, M)}
   {\spconfig{\parconfig{\spconfig{q_1}{\bot}, \ldots , \spconfig{q_n}{\bot}}}{\sigma q} \sptrans{\join}{A} \spconfig{q'}{\sigma}}
\end{mathpar}
where we write $\multiset{q_1, \ldots, q_n}$ for the multiset of states, and
$\trunc{\kappa}{M}$ for the $\kappa$-\emph{truncation} of a multiset $M\in
\mathcal{M}(Q)$, defined as $\trunc{\kappa}{M}(q) = \min(M(q), \kappa)$  for
every $q \in M$.
\end{definition}
 Rules \tcall and \tret are the usual VPA transitions: 
on a \call symbol, the automaton updates its sequential configuration's state and pushes a stack symbol; on a \return symbol, the automaton transitions using the return transition function and pops the stack. 

Rule \teps is the trivial transition for stalled branches. Rules \tfork, \tpar,
and \tjoin handle the parallel regions without commiting to a specific interleaving. 
When \tfork is applied on reading some $\fork_n$, the sequential configuration becomes a parallel configuration with $n$ branch configurations $\spconfig{\parconfig{\spconfig{q'}{\bot}, \ldots , \spconfig{q'}{\bot}}}{\sigma q}$, each initialized at the same state and with an empty stack. The fork state is pushed onto the outer stack, so the automaton can recover it at \join. Rule \tpar advances each branch on its local token $e_i$. If a branch has no symbol to emit while others continue, since its local incremental token is $\epsilon$, it steps by \teps. A completed branch has an empty local stack, meaning it has entirely read the well-matched word. Once all branches have completed, \tjoin aggregates the multiset of all branch states; truncates it at $\kappa$; and collapses the local configurations into a single sequential configuration by applying the join transition to the multiset of terminal branch states.
The truncation step is a technical device: for our compilation, we will
statically select the parameter $\kappa$ large enough so that truncation will
not affect the multiset.

\subsection{Single-step incremental semantics}
The transition relation $\sptrans{e}{\mathcal{A}}$  describes how an SPVPA
reacts to an incremental token, but it does not yet describe how such tokens are
read from an SPNW. For instance, from the word $\APIcall{a}~
\fjword{2}{\APIcall{b}~\APIret{b}, ~\APIcall{c~}\APIret{c}}~ \APIret{a}$, the
monitor first observes $\APIcall{a}$; then $\fork_2$; and then the structured
events $\wordpar{2}{\APIcall{b}, \APIcall{c}}$, and so on. So before we can
define how an SPNW is incrementally processed by an SP-VPA, we need to describe
how an incremental token is read from an SPNW. We formalize this using residual
words, which intuitively represent the remaining SPNW after reading some prefix.

\begin{definition}[Residual words]
We define the set of residual SPNWs, written $Res$, by the grammar:
\[
\begin{aligned}
r ::= {}& \epsilon
      \mid a \cdot r \qquad (a \in \Sigma)\\
      &\mid \fjword{n}{r_1,\ldots,r_n} \cdot r\\
      &\mid \jword{n}{r_1,\ldots,r_n} \cdot r .
\end{aligned}
\]
\end{definition}
The last term represents a parallel block whose opening \fork has already been
read but whose branches have not all completed. Every complete SPNW is a
residual word, \textit{i.e.,} $SPNW \subseteq Res$.

Now, we define how incremental tokens are read from a residual word.
\begin{definition}[Incremental Read]\label{def:read}
For each incremental token $e \in \token$, the relation $\readword{e} \subseteq Res \times Res$ relates a residual word to the residual word that remains after reading $e$.

\begin{mathpar}
  \inferrule*[right=\rcall]{ }{\APIcall{a} \cdot r \readword{\APIcall{a}} r }
  \and
  \inferrule*[right=\rret]{ }{\APIret{a} \cdot r \readword{\APIret{a}} r}
  \and
  \inferrule*[right=\reps]{ }{r \readword{\epsilon} r}
  \and
  \inferrule*[right=\rfork]{ }{\fjword{n}{r_1, \ldots, r_n} \cdot r \readword{\textsf{fork}_n} \jword{n}{r_1, \ldots, r_n} \cdot r}
  \and
    \inferrule*[right=\rpar]
    {r_1 \readword{e_1} r_1' \\ \cdots \\ r_n  \readword{e_n} r_n'}
    {\jword{n}{r_1, \ldots, r_n} \cdot r \readword{\parnew(e_1, \ldots, e_n)} \jword{n}{r_1', \ldots, r_n'} \cdot r}
    \and
  \inferrule*[right=\rjoin]{ }{\jword{n}{\epsilon, \ldots, \epsilon} \cdot r \readword{\textsf{join}} r}
\end{mathpar}
\end{definition}

Rules \rcall and \rret consume the leading call or return symbol; \reps leaves
the residual word unchanged, modeling a branch that does not advance in the
current step. Rule \rfork consumes the opening $\fork_n$ and exposes the pending
parallel block. Rule \rpar performs one read step in each concurrent branch,
possibly $\epsilon$; the lockstep presentation abstracts away concrete
interleavings. Incremental reads continue until all branches have been reduced
to $\epsilon$, meaning they have completed. At that point, \rjoin consumes the
\join symbol and resumes sequential reading.

\begin{figure}
\centering
\begin{subfigure}[t]{0.58\textwidth}
\begin{tikzpicture}[font=\scriptsize, on grid, yscale=.82]

  \node[anchor=west] (r0) at (0,   0mm) {\textsc{Call}};
  \node[anchor=west]       at (1.6cm, 0mm) {$\boxed{\langle\textsf{A}}\ \mathsf{fork}_2\ \parnew_2(\langle\textsf{B}\ \APIret{B}\ \langle\textsf{C}\ \APIret{C})\ \mathsf{join}\ \textsf{A}\rangle$};
  \node[anchor=west]       at (6.2cm, 0mm) {$\to \APIcall{A}$};

  \node[anchor=west] (r1) at (0,  -10mm) {\textsc{Fork}};
  \node[anchor=west]       at (1.6cm,-10mm) {$\textcolor{gray}{\langle\textsf{A}}\ \boxed{\mathsf{fork}_2}\ \parnew_2(\langle\textsf{B}\ \APIret{B}\ \langle\textsf{C}\ \APIret{C})\ \mathsf{join}\ \textsf{A}\rangle$};
  \node[anchor=west]       at (6.2cm,-10mm) {$\to\ \fork_2$};

  \node[anchor=west] (r2a) at (0, -20mm) {\textsc{Par}};
  \node[anchor=west]        at (1.4cm,-20mm) {$\textcolor{gray}{\langle\textsf{A}\ \mathsf{fork}_2}\ \parnew_2(\boxed{\langle\textsf{B}}\ \APIret{B},\ \boxed{\langle\textsf{C}}\ \APIret{C})\ \mathsf{join}\ \textsf{A}\rangle$};
  \node[anchor=west]        at (6.2cm,-20mm) {$\to\ \parnew(\APIcall{B},\ \APIcall{C})$};

  \node[anchor=west] (r2b) at (0, -30mm) {\textsc{Par}};
  \node[anchor=west]        at (1.4cm,-30mm) {$\textcolor{gray}{\langle\textsf{A}\ \mathsf{fork}_2}\ \parnew_2(\textcolor{gray}{\langle\textsf{B}}\ \boxed{\APIret{B}},\ \textcolor{gray}{\langle\textsf{C}}\ \boxed{\APIret{C}})\ \mathsf{join}\ \textsf{A}\rangle$};
  \node[anchor=west]        at (6.2cm,-30mm) {$\to\ \parnew(\APIret{B},\ \APIret{C})$};

  \node[anchor=west] (r3) at (0, -40mm) {\textsc{Join}};
  \node[anchor=west]       at (1.6cm,-40mm) {$\textcolor{gray}{\langle\textsf{A}\ \mathsf{fork}_2\ \parnew_2(\langle\textsf{B}\ \APIret{B}\ \langle\textsf{C}\ \APIret{C})}\ \boxed{\mathsf{join}}\ \textsf{A}\rangle$};
  \node[anchor=west]       at (6.2cm,-40mm) {$\to\ \join$};

  \node[anchor=west] (r4) at (0, -50mm) {\textsc{Ret}};
  \node[anchor=west]       at (1.6cm,-50mm) {$\textcolor{gray}{\langle\textsf{A}\ \mathsf{fork}_2\ \parnew_2(\langle\textsf{B}\ \APIret{B}\ \langle\textsf{C}\ \APIret{C})\ \mathsf{join}}\ \boxed{\textsf{A}\rangle}$};
  \node[anchor=west]       at (6.2cm,-50mm) {$\to\ \APIret{A}$};

  \node[anchor=west] (r5) at (0, -60mm) {};

\end{tikzpicture}
\caption{}
\label{fig:read}
\end{subfigure}%
\hfill
\begin{subfigure}[t]{0.38\textwidth}
\begin{tikzpicture}[font=\scriptsize, on grid, yscale=.82]

  \node[anchor=west] at (0,    8mm) {};
  \node[anchor=center] at (2.8cm,  8mm) {\textit{Configuration}};
  \draw[gray, thin] (0.75cm, 5mm) -- (4.85cm, 5mm);

  \node[anchor=west] at (0.85cm,  0mm) {$(q_0,\ \bot\gamma_a)$};

  \node[anchor=west] at (0.85cm,-10mm) {$\langle \parnew((q_f, \bot)\ (q_f, \bot)\rangle, \ \bot \gamma_a q_0)$};

  \node[anchor=west] at (0.85cm,-20mm) {$\langle \parnew((q_1, \bot \gamma_b)\ (q_2, \bot \gamma_c)\rangle, \ \bot \gamma_a q_s)$};
  \node[anchor=west] at (0.85cm,-30mm) {$\langle \parnew((q_B, \bot)\ (q_C, \bot)\rangle, \ \bot \gamma_a q_s)$};

  \node[anchor=west] at (0.85cm,-40mm) {$(q_{BC}, \ \bot \gamma_a)$};
  \node[anchor=west] at (0.85cm,-50mm) {$(q_{A}, \ \bot)$};

  \node[anchor=west] at (0.85cm,  -60mm) {\textit{(accepted)}};

  \foreach \y in {-5mm,-15mm,-25mm,-35mm,-45mm,-55mm}
    \draw[gray!40, very thin] (0.75cm,\y) -- (4.85cm,\y);

\end{tikzpicture}
\caption{}
\label{fig:run}
\end{subfigure}

\caption{(a)  Incremental read of $\APIcall{A}\ \fjword{2}{\APIcall{B}\ \APIret{B},\ \APIcall{C}\ \APIret{C}}\ \APIret{A}$ 
 and (b) the run of SP-VPA in \cref{fig:spvpa}.}
\label{fig:incremental-read}
\end{figure}

\begin{example}
\Cref{fig:incremental-read} describes the incremental read of the word $\APIcall{A}\;\fjword{2}{\APIcall{B}\;\APIret{B},\;\APIcall{C}\;\APIret{C}}\;\APIret{A}$, which begins by sequentially reading the call to $\APIname{A}$ followed by reading the $\fork$ symbol and entering the parallel block with two branches. There are two applications of \rpar  in the parallel block. In the first step, the first symbols of each branch are read in lockstep followed by the second symbols.  Finally, the \join and final return symbol is read. 
\end{example}

Finally, we define an \textit{incremental step relation} for SP-VPA by combing the incremental read relation with the SP-VPA transition relation. This step relation describes,  for a given SPNW or residual word, both the incremental token exposed in the current step and the corresponding update to the automaton configuration. 

\begin{definition}[Incremental Step]
Given a residual word $r\in Res$, an \textit{incremental step} is a relation $(T, r)\restep{e}(T', r')$ between configurations defined by:
\[
  \inferrule*[]{T \sptrans{e}{A} T' \\ r \readword{e} r'}{(T, r) \restep{e} (T', r')}
\]
\end{definition}
In an incremental step, $\mathcal{A}$ reads the next incremental token $e$ from the residual word $r$ and updates the configuration from $T$ to $T'$, leaving the residual word $r'$.
Finally, we can define our semantics for SP-VPA.

\begin{definition}[SP-VPA Semantics]
A \emph{run} of an SP-VPA $\mathcal{A}$ on a residual word $r$ is defined as the
sequence $(\spconfig{q_0}{\bot},
r)\restep{e_1}\ldots\restep{e_n}(\spconfig{q_f}{\bot}, \epsilon)$ of incremental
steps until the entire word $r$ is read and the residual is $\epsilon$. A run is
\emph{accepting} if $T' = (q_f, \bot)$, where $q_f$ is a final state of the
automaton, otherwise it is \emph{rejecting}.
\end{definition}

\begin{example}
\Cref{fig:incremental-read}(b) illustrates the run of the SP-VPA in \cref{fig:spvpa} on the word $\APIcall{A}~\fjword{2}{\APIcall{B}~\APIret{B},
\APIcall{C}~\APIret{C}}~\APIret{A}$. At each step, the automaton consumes an
incremental token and steps via incremental read and SP-VPA transition rules.
The input word is accepted by the automaton as it arrives at the final state
$q_{A}$.
\end{example}

\subsection{Compilation}
To automatically check whether a SPNW is accepted by a SafePar policy $spe$, we
define a sound compilation procedure from $spe$ to SP-VPA $\mathcal{A}_{spe}$.
We sketch our Thompson-style construction here, and provide details in the
Appendix. 
\begin{enumerate}
\item For $\cs~spe~\rs$, the construction adds a fresh entry state $q_s$, a fresh final state $q_f$, and a fresh stack marker $\gamma$. On reading $\cs$, the automaton pushes $\gamma$ and enters $\mathcal{A}_{spe}$. Once $\mathcal{A}_{spe}$ reaches its final state, the automaton reads the matching $\rs$ and move to $q_f$.
\item For \textit{parallel composition}, $\parnew(spe_1^{m_1}, \ldots, spe_k^{m_k})$, the construction adds a new start state $q_s$ and accepting state $q_f$ to the union of $\mathcal{A}_{spe_i}$. At a fork, each branch is dispatched to the unique sub-automaton determined by its first visible symbol. If the branch begins with a symbol in $\First(spe_i)$ then it is processed by $\mathcal{A}_{spe_i}$. Once a branch has entered $\mathcal{A}_{spe_i}$, its remaining symbols are processed according to this automaton's transitions. At the join, the automaton inspects the multiset of terminal branch states and transitions to the final state if the multiset satisfies the specified multiplicity constraints of the policy. 
\item For \textit{union, concatenation, and Kleene star}, we follow the usual Thompson-style construction.
\end{enumerate}
Finally, we can show that our construction is sound; we detail the proof in the
Appendix.
\begin{theorem}[Soundness]
Given a SafePar policy $spe$ and its automaton $\mathcal{A}_{spe}$, we have:
\[\mathcal{L}(\mathcal{A}_{spe}) = L(spe).\]
\end{theorem}

\section{\safepar Monitor Implementation}\label{sec:impl}

\begin{figure*}[]
  \centering
  \includegraphics[width=0.7\textwidth]{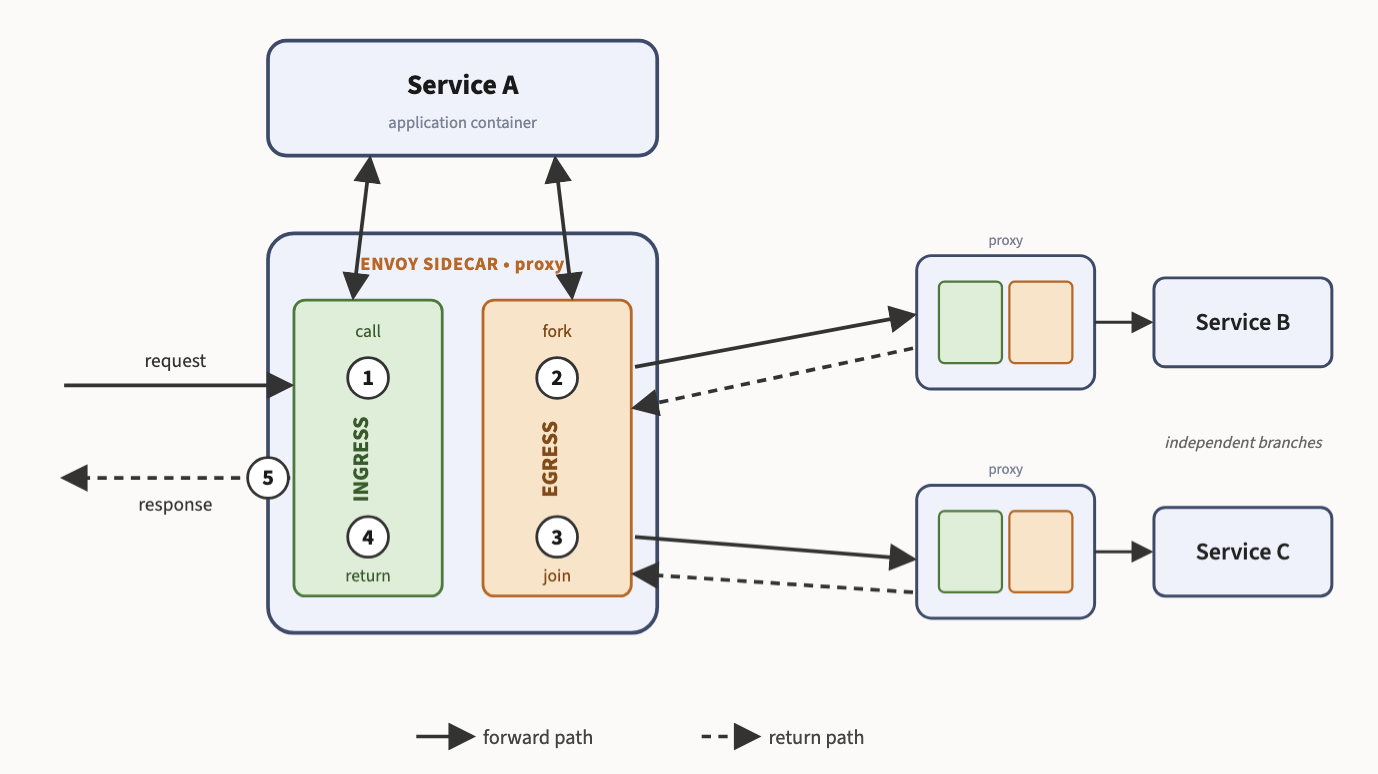}
  \caption{SafePar monitor deployed atop Istio service mesh}
  \label{fig:safet-overview}
\end{figure*}

\newpar{Prototype} We implement a \safepar prototype in $\sim 2$ kLoC that compiles a \safepar policy into an SP-VPA and extracts  a monitor that is deployed as an Envoy WebAssembly filter atop the Istio service mesh layer. The deployment follows the same non-invasive setting as \safetree. Each microservice runs in an independent service container. The service mesh
framework pairs each service container with a sidecar container that
implements an Envoy proxy  \cite{envoydoc} (as shown in \cref{fig:safet-overview}). The proxy intercepts all HTTP
requests and responses corresponding to its service and can read the HTTP
message headers, add/delete/update the
headers, or allow/block HTTP messages. Thus, \safepar does not require any modifications to the  service implementation. The lightweight monitoring metadata, \textit{i.e.,} the current SP-VPA state, is propagated as an  HTTP header. 

In the service mesh, the sidecar proxy at a service handles the call and return events for its own service. 
This design enables SafePar's distributed monitoring, where the monitoring metadata is locally updated, rather than by a centralized monitor.

Each Envoy proxy exposes two filter hooks: \textit{inbound} and \textit{outbound}, depicted in \cref{fig:safet-overview} as green and orange boxes inside the proxy. The \textit{inbound} filter fires when a request arrives at the service and again when the service's response is sent back to its caller. The \textit{outbound} filter fires when the service issues requests to other services and when the corresponding child responses return to the caller. Together, these hooks implement the SP-VPA's call/return and fork/join transitions in a local, per request manner.

\newpar{Inbound call/return monitoring}
When a request arrives at a service, the inbound filter reads the current automaton state from the request header and runs the SP-VPA's call  transition for the service's call symbol. The current state header is updated with the successor state. The stack symbol pushed by this transition is locally saved at the proxy rather than transmitting the entire stack over the network. This works because in our setting, matching call and return events are observed at the same proxy, so the  local value can be looked up when running the return transition.

\newpar{Outbound fork/join monitoring}
Fork and join are  handled analogously at the outbound filter. Before a forked child request exits the proxy, the outbound filter runs the fork transition and updates the state header in the child request. When all child requests return, the final states in their responses are  aggregated and checked against the join predicate to determine next state. 

For simplicity, our prototype logs any policy violation by checking whether, after processing the trace, the SP-VPA is in an accepting configuration or not. The same mechanism can be extended to actively block the request as soon as the monitor reaches a state for which there exists no suffix that can lead to an accepting state.  

\section{Evaluation}\label{sec:eval}
We evaluate SafePar monitor along two dimensions: the header metadata required to carry the monitor's  configuration, and the latency  overhead introduced by monitoring. We structure the evaluation around the following research questions:
\begin{itemize}
\item \textit{RQ1: How much header space is required to carry the configuration metadata?}
\item \textit{RQ2: How much latency overhead is incurred by monitoring?}
\item \textit{RQ3: How does the latency overhead change as we vary the topology scale?}
\end{itemize}
We evaluate SafePar on a suite of policies that cover the main language
constructs. The SafePar monitor is deployed in an Istio-enabled Kubernetes
cluster. We instantiate the policy suite on two Go-based microservice
applications: a hospital workflow and a hotel reservation application
\cite{deathstar} running in the cluster. Each service in the application gets a
proxy and is injected with a WebAssembly SafePar filter. The average number of
nodes in the service tree of both applications is 6 and 4.5, respectively. Since
SafePar runs outside the application, its performance overhead is not affected
by the application's internal implementation. However, its performance does
depend on the policies being checked, which we evaluate. The case studies and
the detailed deployment information is in Appendix \appref{sec:pol}.

\subsection*{Policy metadata}
Since
SafePar enforces policies by propagating the current SP-VPA state in HTTP headers, the memory footprint of the monitoring metadata is determined by the number of monitor states.  \Cref{tbl:eval}'s ``SP-VPA'' columns give the number of states (in \#states), the number of bits needed to encode the current state in a header (in \#bits), and the number of transitions in the monitor (in \#trans). The ``Class'' columns describe the policy structure: \#Name indicates whether the policy is purely sequential (Series), purely parallel (Parallel), or combines sequential and parallel (Series--Par); \#Par reports the number of distinct parallel expressions in the policy; and \#Nesting reports the maximum depth of nesting in the policy. The key result is that across all the policies, the monitoring metadata is less than six bits of HTTP header,  which is small compared with the kilobytes worth of available HTTP header space. \textbf{We observe that policies with more parallel structure and deeper nesting tend to compile to monitors with more states.} The ``Map-reduce'' policy is an exception because, despite its shallow nesting and fewer number of parallel operators, the multiple occurrences of Kleene star operator yields the largest SP-VPA.

\begin{table}[t]

\caption{Evaluation summary for the SafePar policy suite. Policies prefixed with ``Hotel'' are evaluated on the hotel application; the remaining policies are evaluated on the hospital application. For each policy, the number of SP-VPA states fits in a few header bits and monitoring adds only millisecond-scale latency overhead.}
\label{tbl:eval}
\centering
\small
{\begin{tabular}{l ccc l c c c} \toprule & \multicolumn{3}{c}{SP-VPA} & \multicolumn{3}{c}{Class} & Latency \\ \cmidrule(lr){2-4} \cmidrule(lr){5-7} \cmidrule(lr){8-8} \emph{Policy} & \#states & \#bits & \#trans & Name & \#Par & Depth & overhead (ms) \\
\midrule
Scoped order  & 11 & 4 & 10 & Series      & 0 & 2   & 1.11 \\
Bounded-write &  7 & 3 &  7 & Parallel    & 1 & 2    & 1.26\\
Quorum1        & 10 & 4 & 10 & Series--Par & 1 & 2    & 0.35\\
Once-reset    & 15 & 4 & 16 & Series--Par    & 1 & 3    & 0.42\\
Two-phase     & 13 & 4 & 13 & Series--Par & 2 & 2    & 0.53\\
Nested  & 11 & 4 & 12 & Parallel    & 2 & 3    & 0.59\\
Choice        &  9 & 4 &  9 & Series      & 2 & 2                                    & 0.41\\
Agent-phase   & 17 & 5 & 23 & Series--Par & 2 & 3       & 0.73\\
Audit    & 15 & 4 & 17 & Parallel    & 3 & 4       & 0.83\\
Mixed-mult    & 16 & 4 & 16 & Series--Par & 3 & 2   & 0.77\\
Map-reduce    & 26 & 5 & 36 & Series--Par & 2 & 3         & 0.64\\
Quorum2   & 18 & 5 & 21 & Parallel    & 2 & 3                       & 0.56\\
Hotel-Bounded-write &  7 & 3 &  7 & Parallel    & 1 & 2    & 0.435\\
Hotel-Quorum1        & 10 & 4 & 10 & Series--Par & 1 & 2    & 0.42\\
Hotel-Once-reset    & 15 & 4 & 16 & Series--Par    & 1 & 3    & 0.142\\
Hotel-Two-phase     & 13 & 4 & 13 & Series--Par & 2 & 2    & 0.158\\
\bottomrule
\end{tabular}
}
\end{table}

\subsection*{Monitoring overhead}
To evaluate SafePar's  latency overhead, we compare request latency with monitoring enabled against latency without monitoring. We use a workload of 200 concurrent requests to frontend endpoints of both applications in our cluster. For each policy,  we extract and deploy the corresponding Envoy filter and report the average difference between the latency when the application is monitored and when it is not. The ``Overhead'' column in \cref{tbl:eval} reports this monitoring cost in milliseconds.  Observe that across the evaluated policies, the measure overhead is at most 1.5ms. The overhead is also stable across both applications, as expected because SafePar monitors requests at the servicemesh layer and does not depend on the internal implementation of the services. \textbf{To summarize, SafePar can enforce expressive policies with low latency overhead.}

\subsection*{Topology scaling}
We next experimentally explore how the application's API call topology affects monitor performance. We use a synthetic topology generator with varying fanout, concurrency mode, and depths. For each topology, we construct two variants with the same number of API calls. In the synchronous variant, calls are issued sequentially, whereas in the asynchronous variant, all calls at the same level are issued in parallel.  We fix the policy to Quorum1 and run the same overhead experiment from the previous section.

\begin{figure}[t]
    \centering
   \begin{subfigure}[t]{0.32\textwidth}
        \centering
        \includegraphics[height=3cm, width=\textwidth]{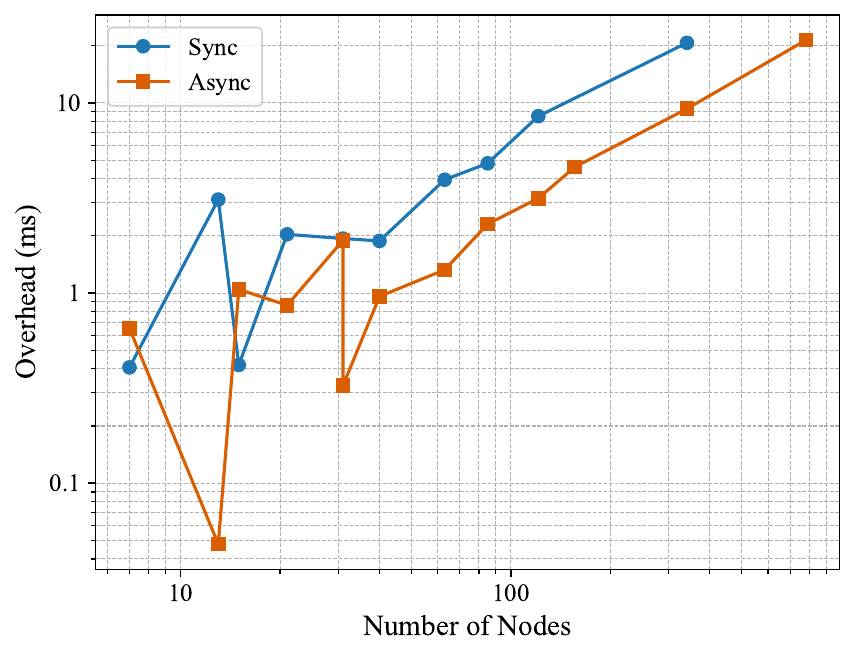}
        \caption{Overhead vs Calls}
        \label{fig:con-sync}
    \end{subfigure}
    \begin{subfigure}[t]{0.32\textwidth}
        \centering
        \includegraphics[height=3cm, width=\textwidth]{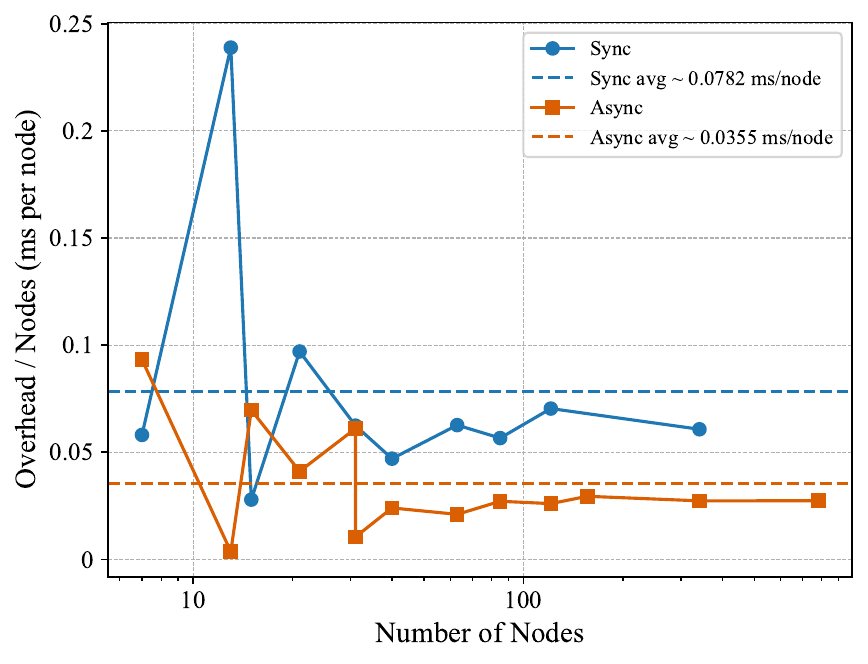}
        \caption{Ratio (Overhead/Calls)}
    \label{fig:ratio-conc-sync}
    \end{subfigure}
    \caption{Latency overhead vs topology scale, measured as number of API calls in the trace.}
    \label{fig:scalability}
\end{figure}

\Cref{fig:con-sync} reports the latency overhead in millisecond on the y-axis
and the topology size measured as number of API calls on the x-axis, both on
logarithmic scale. The overhead grows with the topology size in both the
execution models, but remains below $10$ms for topologies containing
approximately $100$ API calls and below $25$ms even for the largest topologies
considered. Empirical studies \cite{alibaba} report that a common-case microservice topology contains approximately $31$ API calls;  at this scale, SafePar introduces only a few milliseconds of additional latency. 

For topologies of comparable size, the synchronous variant (in blue) generally
incurs higher end-to-end overhead than the asynchronous variant (in orange).
This can be attributed to all monitoring operations being on the request's
critical path in the sequential setting, so the cost of intercepting a request
and updating its headers at each call/return accumulates end to end. In
contrast, SafePar processes asynchronous branches independently and combines
their states only at the join, yielding lower overhead in asynchronous
settings. 

We also report the per-API call overhead for both models of execution in \cref{fig:ratio-conc-sync}. The average per-hop overhead for totally synchronous and totally parallel topology is 0.078 and 0.035, respectively. 
Apart from greater variability for the smallest topologies, where fixed costs constitute a larger fraction of total latency, the per-node overhead remains mostly stable as the topology grows. These results indicate that \textbf{\textit{SafePar's monitoring cost scales approximately linearly with the number of calls.}}


\section{Related Work}
\label{sec:rw}

\newpar{Trace models for concurrent executions}
Classical language-theoretic models often represent 
concurrent executions through their linearizations. 
A run is modeled as a word and the concurrency is accounted 
for by considering the interleavings of concurrent events, 
or more formally a total order between events. 
Mazurkiewicz trace theory refines this view by using a 
partial order semantics and  quotienting the interleaved words 
modulo swaps of independent interleaved events \cite{Mazurkiewicz87,tracetheory}. 
In this model, a concurrent execution denotes an 
equivalence class of linearizations rather than a 
single word recording a specific schedule of events. 
While trace theory offer a very expressive formalism for modeling concurrent
executions, the expressiveness comes at the expense of monitorability,
in that only a small fragment of properties admit efficient
monitors~\cite{Ochmanski85}.
In contrast, in our setting we crucially leverage structured executions,
allowing us to build efficient monitors.

\newpar{Pomsets and series-parallel languages}
Another line of work models executions as partial orders \cite{poset,ba} rather 
than interleaved sequences. Pomsets replace replace words by partial ordered multiset of events, and series-parallel pomsets restrict these partial orders to those generated by sequential and parallel composition. They explicitly capture the series-parallel structure of a concurrent execution and, therefore, are a  natural semantic model for fork-join style concurrent computations \cite{ba}. Branching automata \cite{ba} and pomset automata  \cite{pa1,pa2} recognize languages of series-parallel pomsets using operational semantics that use fork-join transition, in close spirit to our work. The key difference is that these automata models are designed to operate over the completed pomset objects or their algebraic decompositions rather and do not lend themselves to incremental online monitoring over a stream of HTTP call/returns. Moreover, call-return matching is not a primitive part of the pomset model, in the way it is for nested words and visibly pushdown automata.

\newpar{Series-parallel graph automata}
There are several automata models  for concurrent setting, like communicating automata \cite{comaut}, asynchronous automata \cite{Bollig2006,Zielonka1987}. Of the closest interest to us, are series-parallel graph automata \cite{sspga}, which are interpreted over synchronized series-parallel graphs. In this model, series composition represents  causal sequencing, while parallel composition represents fork-join structure. The synchronization edge makes the matching between split and join vertices explicit. The corresponding graph automata also supports join transitions that are guarded by checks on multisets of concurrent branch states. This is the closest prior automata-theoretic analogue to our fork/join transition. However, synchronized series-parallel graph automata are still defined over the whole graph view. Our SP-VPA contribution is to define an automata model that can operate over a stream of word-like inputs with primitive support for enforcing call-return matching. 

\newpar{Runtime verification of concurrent properties}
Prior work has approached verification of concurrent properties from several directions, often with the goal of specifying properties in an interleaving independence manner while keeping the verification procedure tractable. One line of work develops trace-based logic that interpret specifications directly over Mazurkiewicz traces \cite{Leucker2026}; or related partial-order models of concurrency \cite{rvnote,mltl}. A second line of work appears in model checking and testing domain, where the goal  is to reason over partial-order model of executions or to prune redundant interleavings during exploration \cite{globpo, DBLP:conf/lics/AlurPP95,dpor, mcr}
or to infer alternate reorderings from a given execution~\cite{Kini17,FarzanMathur2024,Mathur21}. 
A third line of work focuses on specification-based monitoring and enforcement frameworks for multi-threaded programs such as Enforce MOP and related monitors for generic concurrent systems \cite{enforcemop,survey1,vyrd}. These techniques are often invasive and require instrumentation. SafePar targets a more specialized microservice, where the monitor must support the desired class of concurrent properties, while preserving black-box deployment, non-invasiveness, and distributed execution. As described earlier related work in this domain \cite{safetree,grewalhotnets,whip} do not meet all our requirements.
\section{Conclusion}\label{sec:conclusion}
We present  SafePar, a blackbox and non-invasive distributed runtime monitoring framework for enforcing safety properties over concurrent microservice executions. The key insight behind SafePar is to model   concurrent microservice executions as series-parallel nested words, a formalism we introduce to capture both call-return nesting and fork-join parallelism.  SafePar provides a policy language over SPNWs and  enforces the policies using a series-parallel visibly pushdown automaton.

We see several future directions. First, it would be interesting to extend
SafePar beyond fork-join concurrency to support other asynchronous patterns in
microservice applications, such as callbacks, message queues, and
publish-subscribe communication. Second, SafePar currently focuses on the
structure of the call tree, without reasoning about the parameters  carried by
the requests and responses;  it could also be interesting to extend the
framework to support properties that reason about both the call structure and
the parameter values.

\bibliography{header,safepar}

@STRING{springer = {Springer-Verlag} }

@STRING{lncs     = "Lecture Notes in Computer Science" }

@STRING{pacmpl   = "Proceedings of the {ACM} on Programming Languages" }

@STRING{lipics   = "Leibniz International Proceedings in Informatics" }

@STRING{dagstuhl = "Schloss Dagstuhl--Leibniz Center for Informatics" }

@STRING{ieee     = "IEEE" }

@STRING{aw       = "Addison-Wesley" }

@STRING{popl        = "{ACM} {SIGPLAN--SIGACT} {S}ymposium on {P}rinciples of
                      {P}rogramming {L}anguages ({POPL})" }

@STRING{icfp        = "{ACM} {SIGPLAN} {I}nternational {C}onference on
                      {F}unctional {P}rogramming ({ICFP})" }

@STRING{pldi        = "{ACM SIGPLAN Conference on Programming Language Design
                      and Implementation (PLDI)}" }

@STRING{iclr     = "International Conference on Learning Representations" }

@STRING{stoc04      = stoc # ", Chicago, Illinois"}

@STRING{hotnets     = "{ACM} Workshop on Hot Topics in Networks ({HotNets})" }

@STRING{hotnets23   = hotcloud # ", Cambridge, Massachusetts"}

@STRING{concur    = {International Conference on Concurrency Theory (CONCUR)} }

@STRING{icalp     = {International Colloquium on Automata, Languages and
                    Programming (ICALP)} }

@STRING{issta     = {{ACM} {SIGSOFT} International Symposium on Software Testing and Analysis {(ISSTA)}} }

@STRING{rv        = {International Conference on Runtime Verification (RV)} }

@STRING{asplos19    = asplos # ", Providence, Rhode Island" }

@preamble{"\newcommand{\SortNoop}[1]{}"}

@article{unamb,
author = {Br\"{u}ggemann-Klein, Anne and Wood, Derick},
title = {One-unambiguous regular languages},
year = {1998},
issue_date = {May 2, 1998},
publisher = {Academic Press, Inc.},
address = {USA},
volume = {142},
number = {2},
issn = {0890-5401},
url = {https://doi.org/10.1006/inco.1997.2695},
doi = {10.1006/inco.1997.2695},
journal = {Inf. Comput.},
month = may,
pages = {182–206},
numpages = {25}
}

@article{vyrd,
author = {Elmas, Tayfun and Tasiran, Serdar and Qadeer, Shaz},
title = {VYRD: verifYing concurrent programs by runtime refinement-violation detection},
year = {2005},
issue_date = {June 2005},
publisher = {Association for Computing Machinery},
address = {New York, NY, USA},
volume = {40},
number = {6},
issn = {0362-1340},
url = {https://doi.org/10.1145/1064978.1065015},
doi = {10.1145/1064978.1065015},
journal = {SIGPLAN Not.},
month = jun,
pages = {27–37},
numpages = {11}
}

@InProceedings{ba,
author="Lodaya, K.
and Weil, P.",
editor="Morvan, Michel
and Meinel, Christoph
and Krob, Daniel",
title="Series-parallel posets: Algebra, automata and languages",
booktitle="STACS 98",
year="1998",
publisher="Springer Berlin Heidelberg",
address="Berlin, Heidelberg",
pages="555--565",
isbn="978-3-540-69705-3"
}

@inproceedings{poset,
  author       = {Dietrich Kuske},
  editor       = {Ugo Montanari and
                  Jos{\'{e}} D. P. Rolim and
                  Emo Welzl},
  title        = {Infinite Series-Parallel Posets: Logic and Languages},
  booktitle    = {Automata, Languages and Programming, 27th International Colloquium,
                  {ICALP} 2000, Geneva, Switzerland, July 9-15, 2000, Proceedings},
  series       = {Lecture Notes in Computer Science},
  volume       = {1853},
  pages        = {648--662},
  publisher    = {Springer},
  year         = {2000},
  url          = {https://doi.org/10.1007/3-540-45022-X\_55},
  doi          = {10.1007/3-540-45022-X\_55},
  bibsource    = {dblp computer science bibliography, https://dblp.org}
}

@Inbook{Leucker2026,
author="Leucker, Martin",
editor="Bertrand, Nathalie
and Dubslaff, Clemens
and Kl{\"u}ppelholz, Sascha",
title="A Note on Runtime Verification of Concurrent Systems",
bookTitle="Principles of Formal Quantitative Analysis: Essays Dedicated to Christel Baier on the Occasion of Her 60th Birthday",
year="2026",
publisher="Springer Nature Switzerland",
address="Cham",
pages="253--265",
isbn="978-3-031-97439-7",
doi="10.1007/978-3-031-97439-7_12",
url="https://doi.org/10.1007/978-3-031-97439-7_12"
}

@inproceedings{survey1,
  author       = {Yoriyuki Yamagata and
                  Cyrille Artho and
                  Masami Hagiya and
                  Jun Inoue and
                  Lei Ma and
                  Yoshinori Tanabe and
                  Mitsuharu Yamamoto},
  editor       = {Yli{\`{e}}s Falcone and
                  C{\'{e}}sar S{\'{a}}nchez},
  title        = {Runtime Monitoring for Concurrent Systems},
  booktitle    = {Runtime Verification - 16th International Conference, {RV} 2016, Madrid,
                  Spain, September 23-30, 2016, Proceedings},
  series       = {Lecture Notes in Computer Science},
  volume       = {10012},
  pages        = {386--403},
  publisher    = {Springer},
  year         = {2016},
  url          = {https://doi.org/10.1007/978-3-319-46982-9\_24},
  doi          = {10.1007/978-3-319-46982-9\_24},
  bibsource    = {dblp computer science bibliography, https://dblp.org}
}

@inproceedings{enforcemop,
author = {Luo, Qingzhou and Ro\c{s}u, Grigore},
title = {EnforceMOP: a runtime property enforcement system for multithreaded programs},
year = {2013},
isbn = {9781450321594},
publisher = {Association for Computing Machinery},
address = {New York, NY, USA},
url = {https://doi.org/10.1145/2483760.2483766},
doi = {10.1145/2483760.2483766},
booktitle = {Proceedings of the 2013 International Symposium on Software Testing and Analysis},
pages = {156–166},
numpages = {11},
location = {Lugano, Switzerland},
series = {ISSTA 2013}
}

@article{mcr,
author = {Huang, Jeff},
title = {Stateless model checking concurrent programs with maximal causality reduction},
year = {2015},
issue_date = {June 2015},
publisher = {Association for Computing Machinery},
address = {New York, NY, USA},
volume = {50},
number = {6},
issn = {0362-1340},
url = {https://doi.org/10.1145/2813885.2737975},
doi = {10.1145/2813885.2737975},
journal = {SIGPLAN Not.},
month = jun,
pages = {165–174},
numpages = {10}
}

@article{dpor,
author = {Flanagan, Cormac and Godefroid, Patrice},
title = {Dynamic partial-order reduction for model checking software},
year = {2005},
issue_date = {January 2005},
publisher = {Association for Computing Machinery},
address = {New York, NY, USA},
volume = {40},
number = {1},
issn = {0362-1340},
url = {https://doi.org/10.1145/1047659.1040315},
doi = {10.1145/1047659.1040315},
journal = {SIGPLAN Not.},
month = jan,
pages = {110–121},
numpages = {12}
}

@inproceedings{DBLP:conf/lics/AlurPP95,
  author       = {Rajeev Alur and
                  Doron A. Peled and
                  Wojciech Penczek},
  title        = {Model-Checking of Causality Properties},
  booktitle    = {Proceedings, 10th Annual {IEEE} Symposium on Logic in Computer Science,
                  San Diego, California, USA, June 26-29, 1995},
  pages        = {90--100},
  publisher    = {{IEEE} Computer Society},
  year         = {1995},
  url          = {https://doi.org/10.1109/LICS.1995.523247},
  doi          = {10.1109/LICS.1995.523247},
  bibsource    = {dblp computer science bibliography, https://dblp.org}
}

@inproceedings{globpo,
  author       = {Rajeev Alur and
                  Kenneth L. McMillan and
                  Doron A. Peled},
  editor       = {Kim Guldstrand Larsen and
                  Sven Skyum and
                  Glynn Winskel},
  title        = {Deciding Global Partial-Order Properties},
  booktitle    = {Automata, Languages and Programming, 25th International Colloquium,
                  ICALP'98, Aalborg, Denmark, July 13-17, 1998, Proceedings},
  series       = {Lecture Notes in Computer Science},
  volume       = {1443},
  pages        = {41--52},
  publisher    = {Springer},
  year         = {1998},
  url          = {https://doi.org/10.1007/BFb0055039},
  doi          = {10.1007/BFB0055039},
  bibsource    = {dblp computer science bibliography, https://dblp.org}
}

@Inbook{rvnote,
author="Leucker, Martin",
editor="Bertrand, Nathalie
and Dubslaff, Clemens
and Kl{\"u}ppelholz, Sascha",
title="A Note on Runtime Verification of Concurrent Systems",
bookTitle="Principles of Formal Quantitative Analysis: Essays Dedicated to Christel Baier on the Occasion of Her 60th Birthday",
year="2026",
publisher="Springer Nature Switzerland",
address="Cham",
pages="253--265",
isbn="978-3-031-97439-7",
doi="10.1007/978-3-031-97439-7_12",
url="https://doi.org/10.1007/978-3-031-97439-7_12"
}

@inproceedings{mltl,
author = {Mukund, Madhavan and Thiagarajan, P. S.},
title = {Linear Time Temporal Logics over Mazurkiewicz Traces},
year = {1996},
isbn = {3540615504},
publisher = {Springer-Verlag},
address = {Berlin, Heidelberg},
booktitle = {Proceedings of the 21st International Symposium on Mathematical Foundations of Computer Science},
pages = {62–92},
numpages = {31},
series = {MFCS '96}
}

@inproceedings{pa2,
  author       = {Tobias Kapp{\'{e}} and
                  Paul Brunet and
                  Bas Luttik and
                  Alexandra Silva and
                  Fabio Zanasi},
  editor       = {Roland Meyer and
                  Uwe Nestmann},
  title        = {Brzozowski Goes Concurrent - {A} Kleene Theorem for Pomset Languages},
  booktitle    = {28th International Conference on Concurrency Theory, {CONCUR} 2017,
                  Berlin, Germany, September 5-8, 2017},
  series       = {LIPIcs},
  volume       = {85},
  pages        = {25:1--25:16},
  publisher    = {Schloss Dagstuhl - Leibniz-Zentrum f{\"{u}}r Informatik},
  year         = {2017},
  url          = {https://doi.org/10.4230/LIPIcs.CONCUR.2017.25},
  doi          = {10.4230/LIPICS.CONCUR.2017.25},
  bibsource    = {dblp computer science bibliography, https://dblp.org}
}

@article{pa1,
  author       = {Tobias Kapp{\'{e}} and
                  Paul Brunet and
                  Bas Luttik and
                  Alexandra Silva and
                  Fabio Zanasi},
  title        = {On Series-Parallel Pomset Languages: Rationality, Context-Freeness
                  and Automata},
  journal      = {CoRR},
  volume       = {abs/1812.03058},
  year         = {2018},
  url          = {http://arxiv.org/abs/1812.03058},
  eprinttype   = {arXiv},
  eprint       = {1812.03058},
  bibsource    = {dblp computer science bibliography, https://dblp.org}
}

@article{sspga,
  author       = {Rajeev Alur and
                  Caleb Stanford and
                  Christopher Watson},
  title        = {A Robust Theory of Series Parallel Graphs},
  journal      = {Proc. {ACM} Program. Lang.},
  volume       = {7},
  number       = {{POPL}},
  pages        = {1058--1088},
  year         = {2023},
  url          = {https://doi.org/10.1145/3571230},
  doi          = {10.1145/3571230},
  bibsource    = {dblp computer science bibliography, https://dblp.org}
}

@incollection{comaut,
  author       = {Dietrich Kuske and
                  Anca Muscholl},
  editor       = {Jean{-}{\'{E}}ric Pin},
  title        = {Communicating automata},
  booktitle    = {Handbook of Automata Theory},
  pages        = {1147--1188},
  publisher    = {European Mathematical Society Publishing House, Z{\"{u}}rich,
                  Switzerland},
  year         = {2021},
  url          = {https://doi.org/10.4171/Automata-2/9},
  doi          = {10.4171/AUTOMATA-2/9},
  bibsource    = {dblp computer science bibliography, https://dblp.org}
}

@Inbook{Bollig2006,
title="Mazurkiewicz Traces and Asynchronous Automata",
bookTitle="Formal Models of Communicating Systems: Languages, Automata, and Monadic Second-Order Logic",
year="2006",
publisher="Springer Berlin Heidelberg",
address="Berlin, Heidelberg",
pages="77--90",
isbn="978-3-540-32923-7",
doi="10.1007/3-540-32923-4_6",
url="https://doi.org/10.1007/3-540-32923-4_6"
}

@book{tracetheory,
  editor    = {Volker Diekert and
               Grzegorz Rozenberg},
  title     = {The Book of Traces},
  publisher = {World Scientific},
  year      = {1995},
  opt-url       = {https://doi.org/10.1142/2563},
  opt-doi       = {10.1142/2563},
  isbn      = {978-981-02-2058-7},
  opt-timestamp = {Mon, 22 Jul 2019 19:53:40 +0200},
  opt-biburl    = {https://dblp.org/rec/bib/books/ws/95/DR1995},
  opt-bibsource = {dblp computer science bibliography, https://dblp.org}
}

@article{Ochmanski85,
  author       = {Edward Ochma{\'n}ski},
  title        = {Regular behaviour of concurrent systems},
  journal      = {Bull. {EATCS}},
  volume       = {27},
  pages        = {56--67},
  year         = {1985},
  bibsource    = {dblp computer science bibliography, https://dblp.org}
}

@article{Zielonka1987,
  author  = {Zielonka, Wies{\l}aw},
  title   = {Notes on Finite Asynchronous Automata},
  journal = {RAIRO -- Theoretical Informatics and Applications},
  volume  = {21},
  number  = {2},
  pages   = {99--135},
  year    = {1987},
  doi     = {10.1051/ita/1987210200991},
}

@misc{iclr,
  author = {ICLR},
  title = {{ICLR 2026 Response to Security Incident}},
  HowPublished = {\url{https://blog.iclr.cc/2025/12/03/iclr-2026-response-to-security-incident/
}},
  month = Jul,
  year = 2026,
}

@article{bugmicroservice,
author = {Zhou, Xiang and Peng, Xin and Xie, Tao and Sun, Jun and Ji, Chao and Li, Wenhai and Ding, Dan},
title = {Fault Analysis and Debugging of Microservice Systems: Industrial Survey, Benchmark System, and Empirical Study},
year = {2021},
issue_date = {Feb. 2021},
publisher = {IEEE Press},
volume = {47},
number = {2},
issn = {0098-5589},
url = {https://doi.org/10.1109/TSE.2018.2887384},
doi = {10.1109/TSE.2018.2887384},
journal = {IEEE Trans. Softw. Eng.},
month = feb,
pages = {243–260},
numpages = {18}
}

@article{safetree,
author = {Grewal, Karuna and Godfrey, Brighten and Hsu, Justin},
title = {SafeTree: Expressive Tree Policies for Microservices},
year = {2025},
issue_date = {October 2025},
publisher = {Association for Computing Machinery},
address = {New York, NY, USA},
volume = {9},
number = {OOPSLA2},
url = {https://doi.org/10.1145/3763127},
doi = {10.1145/3763127},
journal = {Proc. ACM Program. Lang.},
month = oct,
articleno = {349},
numpages = {27}
}

@misc{envoydoc,
  author = {Envoy},
  title = {{Envoy Proxy}},
  HowPublished = {\url{https://www.envoyproxy.io/}},
  month = nov,
  year = 2024,
  Note = {Accessed: 2024-11-04}
}

@article{whip,
author = {Waye, Lucas and Chong, Stephen and Dimoulas, Christos},
title = {Whip: higher-order contracts for modern services},
year = {2017},
issue_date = {September 2017},
publisher = {Association for Computing Machinery},
address = {New York, NY, USA},
volume = {1},
number = {ICFP},
url = {https://doi.org/10.1145/3110280},
doi = {10.1145/3110280},
journal = pacmpl,
month = aug,
articleno = {36}
}

@misc{istio,
  author = {Istio},
  title = {{Istio}},
  HowPublished = {\url{https://istio.io/}},
  month = nov,
  year = 2024,
  Note = {Accessed: 2024-11-04}
}

@misc{servicemesh,
  author = {Istio},
  title = {{The Istio service mesh}},
  HowPublished = {\url{
https://istio.io/latest/about/service-mesh/}},
  month = nov,
  year = 2024,
  Note = {Accessed: 2024-11-06}
}

@inproceedings{deathstar,
author = {Gan, Yu and Zhang, Yanqi and Cheng, Dailun and Shetty, Ankitha and Rathi, Priyal and Katarki, Nayan and Bruno, Ariana and Hu, Justin and Ritchken, Brian and Jackson, Brendon and Hu, Kelvin and Pancholi, Meghna and He, Yuan and Clancy, Brett and Colen, Chris and Wen, Fukang and Leung, Catherine and Wang, Siyuan and Zaruvinsky, Leon and Espinosa, Mateo and Lin, Rick and Liu, Zhongling and Padilla, Jake and Delimitrou, Christina},
title = {An Open-Source Benchmark Suite for Microservices and Their Hardware-Software Implications for Cloud \& Edge Systems},
year = {2019},
isbn = {9781450362405},
publisher = {Association for Computing Machinery},
address = {New York, NY, USA},
url = {https://doi.org/10.1145/3297858.3304013},
doi = {10.1145/3297858.3304013},
booktitle = asplos19,
pages = {3--18},
}

@misc{gdpr,
  author = {Proton AG},
  title = {{Complete guide to GDPR compliance}},
  HowPublished = {\url{https://gdpr.eu/}},
  month = nov,
  year = 2024,
  Note = {Accessed: 2024-11-04}
}

@inproceedings{vpa,
  author       = {Rajeev Alur and
                  P. Madhusudan},
  title        = {Visibly pushdown languages},
  booktitle    = stoc04,
  pages        = {202--211},
  publisher    = {{ACM}},
  year         = {2004},
  url          = {https://doi.org/10.1145/1007352.1007390},
  doi          = {10.1145/1007352.1007390},
  bibsource    = {dblp computer science bibliography, https://dblp.org}
}

@inproceedings{nestedwords,
  author       = {Rajeev Alur and
                  P. Madhusudan},
  title        = {Adding Nesting Structure to Words},
  booktitle    = {International Conference on Developments in Language Theory
                  ({DLT}), Santa Barbara, California},
  series       = lncs,
  volume       = {4036},
  pages        = {1--13},
  publisher    = springer,
  year         = {2006},
  url          = {https://doi.org/10.1007/11779148\_1},
  doi          = {10.1007/11779148\_1},
  bibsource    = {dblp computer science bibliography, https://dblp.org}
}

@inproceedings{grewalhotnets,
  author       = {Karuna Grewal and
                  Philip Brighten Godfrey and
                  Justin Hsu},
  title        = {Expressive Policies For Microservice Networks},
  booktitle    = hotnets23,
  pages        = {280--286},
  publisher    = {{ACM}},
  year         = {2023},
  url          = {https://doi.org/10.1145/3626111.3628181},
  doi          = {10.1145/3626111.3628181},
  bibsource    = {dblp computer science bibliography, https://dblp.org}
}

@inproceedings{ashok21servicemeshes,
  author    = {Sachin Ashok and P. Brighten Godfrey and Radhika Mittal},
  title     = {Leveraging Service Meshes as a New Network Layer},
  booktitle = {Twentieth ACM Workshop on Hot Topics in Networks (HotNets)},
  month     = {November},
  year      = {2021},
}

@inproceedings{alibaba,
author = {Luo, Shutian and Xu, Huanle and Lu, Chengzhi and Ye, Kejiang and Xu, Guoyao and Zhang, Liping and Ding, Yu and He, Jian and Xu, Chengzhong},
title = {Characterizing Microservice Dependency and Performance: Alibaba Trace Analysis},
year = {2021},
isbn = {9781450386388},
publisher = {Association for Computing Machinery},
address = {New York, NY, USA},
url = {https://doi.org/10.1145/3472883.3487003},
doi = {10.1145/3472883.3487003},
booktitle = {{ACM} Symposium on Cloud Computing {(SoCC}), Seattle, Washington},
pages = {412--426},
}

@Inbook{microservice,
author="Dragoni, Nicola
and Giallorenzo, Saverio
and Lafuente, Alberto Lluch
and Mazzara, Manuel
and Montesi, Fabrizio
and Mustafin, Ruslan
and Safina, Larisa",
editor="Mazzara, Manuel
and Meyer, Bertrand",
title="Microservices: Yesterday, Today, and Tomorrow",
bookTitle="Present and Ulterior Software Engineering",
year="2017",
publisher="Springer International Publishing",
address="Cham",
pages="195--216",
isbn="978-3-319-67425-4",
doi="10.1007/978-3-319-67425-4_12",
url="https://doi.org/10.1007/978-3-319-67425-4_12"
}

@inproceedings{Mazurkiewicz87,
  author = {Mazurkiewicz, A},
  title = {Trace Theory},
  booktitle = {Advances in Petri Nets 1986, Part II on Petri Nets: Applications and Relationships to Other Models of Concurrency},
  year = {1987},
  pages = {279--324},
  publisher = {Springer-Verlag New York, Inc.}
}

@inproceedings{Kini17,
author = {Kini, Dileep and Mathur, Umang and Viswanathan, Mahesh},
title = {Dynamic Race Prediction in Linear Time},
booktitle = {Proceedings of the 38th ACM SIGPLAN Conference on Programming Language Design and Implementation},
series = {PLDI 2017},
year = {2017},
isbn = {978-1-4503-4988-8},
location = {Barcelona, Spain},
pages = {157--170},
numpages = {14},
url = {http://doi.acm.org/10.1145/3062341.3062374},
doi = {10.1145/3062341.3062374},
acmid = {3062374},
publisher = {ACM},
address = {New York, NY, USA},
}

@article{Mathur21,
author = {Mathur, Umang and Pavlogiannis, Andreas and Viswanathan, Mahesh},
title = {Optimal Prediction of Synchronization-Preserving Races},
year = {2021},
issue_date = {January 2021},
publisher = {Association for Computing Machinery},
address = {New York, NY, USA},
volume = {5},
number = {POPL},
url = {https://doi.org/10.1145/3434317},
doi = {10.1145/3434317},
journal = {Proc. ACM Program. Lang.},
month = jan,
articleno = {36},
numpages = {29}
}

@article{FarzanMathur2024,
author = {Farzan, Azadeh and Mathur, Umang},
title = {Coarser Equivalences for Causal Concurrency},
year = {2024},
issue_date = {January 2024},
publisher = {Association for Computing Machinery},
address = {New York, NY, USA},
volume = {8},
number = {POPL},
url = {https://doi.org/10.1145/3632873},
doi = {10.1145/3632873},
journal = {Proc. ACM Program. Lang.},
month = jan,
articleno = {31},
numpages = {31}
}

\newpage
\ifappendix
\appendix
\fi

\end{document}